\documentclass[a4paper,11pt]{article}

\usepackage{jinstpub} 

\usepackage[version=4]{mhchem}
\usepackage{lipsum}
\setlipsum{%
  par-before = \begingroup\color{gray},
  par-after = \endgroup
}
\usepackage{siunitx}
\usepackage{subfig}
\usepackage{graphicx}

\usepackage{booktabs, multirow} 
\usepackage{soul}
\usepackage{changepage,threeparttable} 
\usepackage{orcidlink}

\title{\boldmath Discrete Wavelet Transform for Serial X-ray Crystallography Image Segmentation}

\author{$^{1}$D. Doering\orcidlink{0000-0001-5182-9044},}
\author{$^{1}$N. Claret,}
\author{$^{1}$G. Paulino,}
\author{$^{1}$L. Scomparin\orcidlink{0000-0002-1773-0769},}
\author{$^{1}$F. Poitevin,}
\author{$^{4}$E. Darve,}
\author{$^{1}$C. Hansson,}
\author{$^{1}$J. Russell,}
\author{$^{1}$A. Dave,}
\author{$^{1}$L. Rota,}
\author{$^{2,3}$A. Miceli\orcidlink{0000-0001-5994-5402}}
\author{$^{1}$R. Herbst,}
\author{and $^{1}$A. Dragone}

\affiliation{$^1$SLAC National Accelerator Laboratory, Menlo Park, CA, U.S.A.}
\affiliation{$^2$Argonne National Laboratory, Lemont, IL, U.S.A.}
\affiliation{$^3$University of Chicago, Chicago, IL, U.S.A.}
\affiliation{$^4$Institute for Computational
and Mathematical Engineering,
Stanford, USA}

\emailAdd{ddoering@slac.stanford.edu}

\abstract{
Upcoming LCLS-II/II-HE operation at repetition rates approaching 1~MHz demands on-detector data reduction to manage the resulting data volumes. We present a 2D discrete wavelet transform (DWT) pre-processing algorithm that segments background scatter from crystal diffraction in serial crystallography images, enabling early data analysis and, when combined with peak finding, lossy compression by transmitting only the identified diffraction peaks. The method zeroes the approximation (LL) coefficients of a multi-level Haar wavelet decomposition and reconstructs from detail subbands only, exploiting the natural separation of smooth background and sharp Bragg peaks in the wavelet domain. Evaluated on 100 simulated nanoBragg frames with known ground truth, the pipeline achieves $F1 \approx 0.96$ at four decomposition levels ($J = 4$), substantially outperforming the established peakfinder8 algorithm ($F1 \approx 0.37$) in both precision ($P \approx 1.00$ vs.\ $0.94$) and recall ($R \approx 0.92$ vs.\ $0.24$). A comparison of 12~wavelet families confirms that Haar is optimal for Bragg-peak detection due to its minimal filter support. Downstream crystallographic analysis performed on real ePix10kA data shows that CC* and $R_\mathrm{split}$ converge at $J = 4$ and track the unprocessed baseline through the practical resolution limit. Under added noise exceeding $\sim$50~ADU, the current pipeline's precision degrades significantly more than that of the pf8 algorithm, exposing a limitation of the proposed strategy. We also demonstrate an FPGA implementation of the DWT filters on an Alveo U200 at 200~MHz, with a projected resource footprint compatible with integration into the upcoming ePix~UHR firmware and a path to on-detector ASIC implementation in SparkPix detector family.
}


\begin{document}
\maketitle
\flushbottom


\section{Introduction}
\label{sec:introduction}
\begin{figure}[htp!]
  \centering
  \centerline{\includegraphics[width=16.5cm]{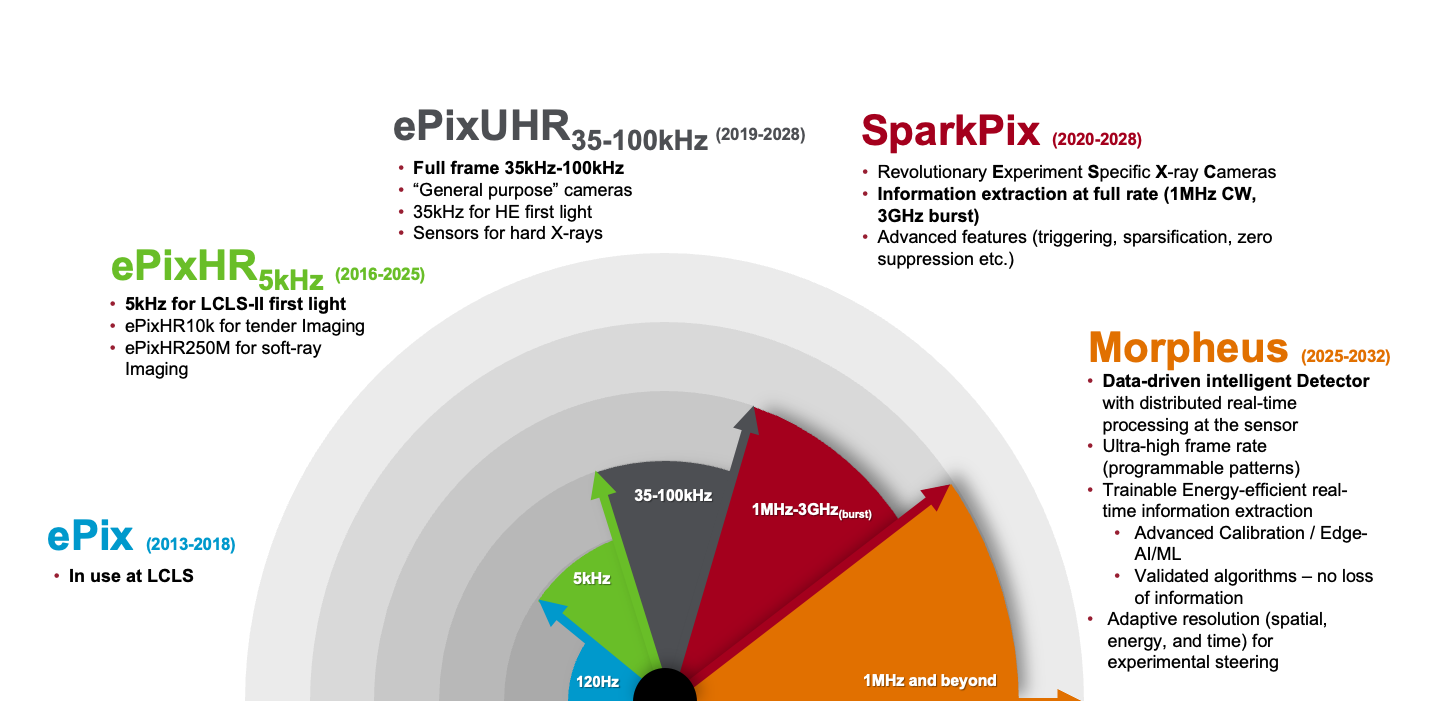}}
  \caption{SLAC X-ray detector families.}
  \label{fig:ePix}
\end{figure}

LCLS-II \cite{schoenlein2016lcls} is a free-electron-laser (FEL) X-ray light source that began operations at SLAC National Laboratory in 2020. Recent upgrades allow the machine to operate at rates approaching 1 MHz and expand its X-ray energy range from the soft to the hard X-ray regime. These changes directly affect detector requirements: detectors in the experimental hutches must run at the accelerator rate to maintain shot-by-shot imaging. The detector development has addressed this rate increase in a phased, systematic manner. Figure \ref{fig:ePix} shows the detector family, where each member marks a key milestone in achievable frame rate. Full-frame readout is used up to ePix UHR, which supports 35,000–100,000 frames per second (fps) \cite{Sandberg2025, king2023characterization}. Beyond 100,000 fps, SparkPix \cite{rota2024sparkpix, markovic2023sparkpix} and its variants introduce data reduction at the ASIC level, for example, transmitting only pixels whose signal exceeds a programmable threshold while suppressing lower-energy pixels.

\begin{figure}[htp!]
  \centering
  \centerline{\includegraphics[width=10.5cm]{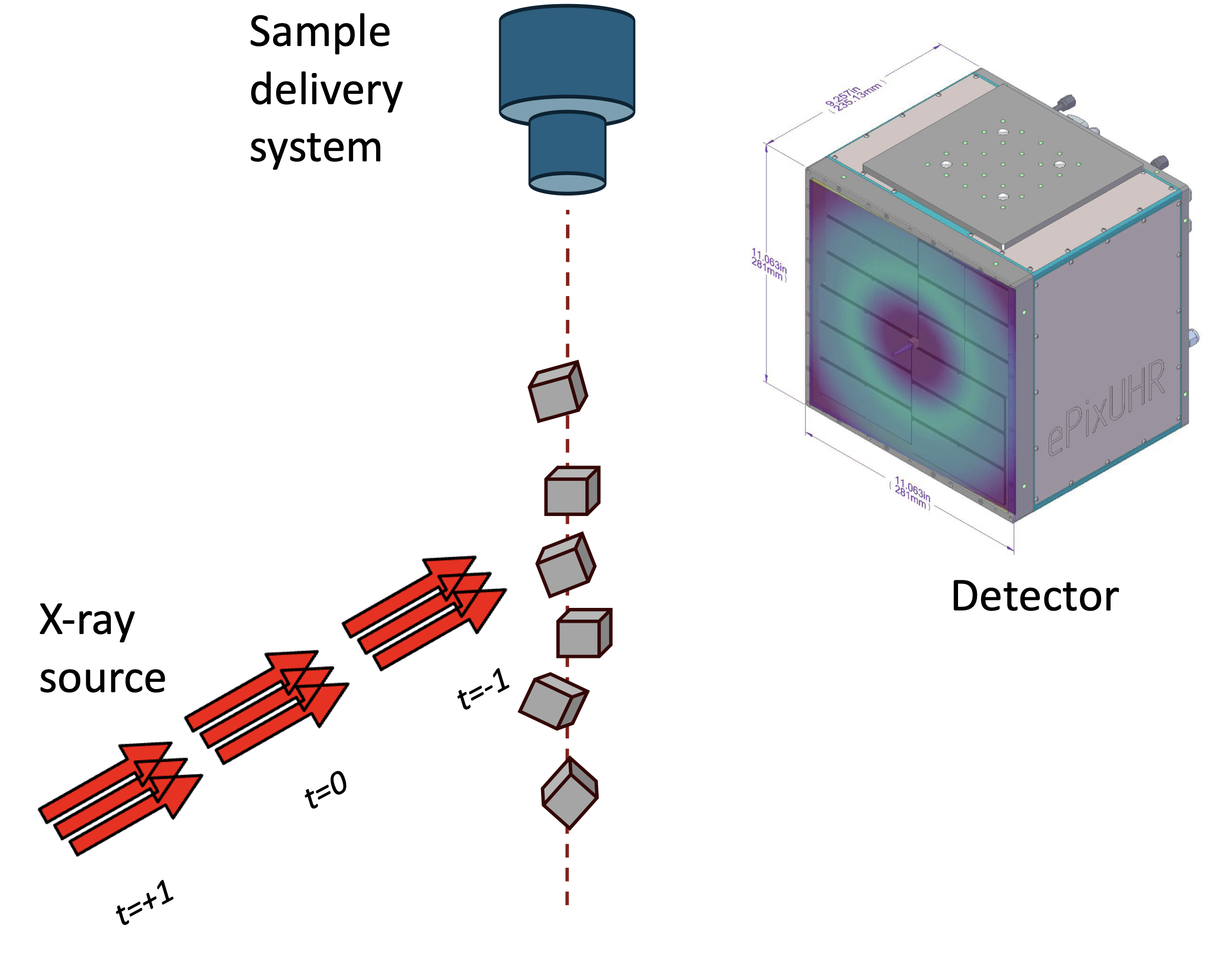}}
  \caption{Serial Crystallography experimental setup.}
  \label{fig:serial_crystallography_exp_setup}
\end{figure}

In full-frame modes such as ePix UHR, all pixel data can be transmitted; however, at 35,000–100,000 fps the data volume remains substantial, prompting several approaches to manage the deluge \cite{rasheedi202528}. The algorithms for data reduction tend to be application specific. One application that is envisioned for the ePixUHR detector is serial crystallography. Figure \ref{fig:serial_crystallography_exp_setup} illustrates a typical serial crystallography experiment: an X-ray beam interacts with crystals under analysis, and, under the intense pulses, a “diffract-before-destroy” approach is used, in which crystals are embedded in a liquid and delivered as a jet so that each particle interacts with the beam only once. The resulting diffraction images contain both background (water and air scatter) and signal (crystal diffraction).

In the serial crystallography example, if one can segment the background signal from diffraction images, all the information from the background can be eliminated simplifying the data processing algorithms and potentially enabling hardware compatible algorithms to implement image classification, image compression or other techniques for data reduction. In this paper, we present a real-time pre-processing method based on the Discrete Wavelet Transform (DWT) that performs the segmentation of background and diffraction signals for serial crystallography. We also discuss how this pre-processing can be ported to the available processing units (i.e. ASIC, FPGAs beyond CPU and GPU).

The remainder of this paper is organized as follows. Section~\ref{sec:related_work} reviews related work on background subtraction, deep learning peak detection, and on-detector processing. Section~\ref{sec:methods} presents the methods: the DWT and the segmentation strategy it enables (Section~\ref{sec:dwt}), together with the crystallographic analysis pipeline used to evaluate data quality (Section~\ref{sec:crystallographic_pipeline}). Section~\ref{sec:results} presents qualitative segmentation results for real images taken at LCLS and quantitative segmentation results for simulated images. Section~\ref{sec:hw_implementation} discusses how the algorithm can be ported to hardware. Finally, Section~\ref{sec:Summary} summarizes the paper and discusses future work.

\section{Related Work}
\label{sec:related_work}

\paragraph{Background subtraction in serial crystallography.}
The dominant strategy for background removal in serial crystallography is radial averaging: a spherically symmetric background model is estimated from pixel intensities at a given resolution shell and subtracted from each frame. This approach is simple and computationally cheap, but it assumes the background is radially symmetric, which breaks down when the scattering medium (liquid jet, membrane, cell) introduces anisotropic features. The Cheetah framework \cite{Barty2014} popularised this idea as an online hit-finding step; its peakfinder8 algorithm refines it by computing the background locally and adaptively, retaining only peaks whose local
SNR exceeds a threshold. Hadian-Jazi et al.\ \cite{HadianJazi2021} replace this heuristic with a robust-statistics framework, modelling per-resolution-ring intensities as a Poisson distribution and using the median as an outlier-insensitive background estimate; the method requires fewer tunable parameters than peakfinder8 and improves CC and $R_\text{split}$ at default settings. Kieffer et al.\ \cite{Kieffer2025} work in azimuthal space to separate the isotropic amorphous background from Bragg-peak outliers, demonstrating real-time operation at 925\,Hz on a Jungfrau 4M detector at ESRF and enabling both sparse compression and peak picking. Despite these advances, all four methods rely on the assumption of a radially symmetric
background to separate signal from noise. Our work takes a different approach: the 2D-DWT
decomposition operates directly in pixel space and makes no assumption about background
isotropy, providing a segmentation mechanism that is agnostic to the radial symmetry of the
scattering environment.

\paragraph{Deep learning approaches.}
Neural-network methods have been explored to improve both peak detection accuracy and background separation. BraggNN \cite{BraggNN_Liu2022} replaces traditional peakfinder8 with a lightweight convolutional neural network (CNN) that operates on small image patches around candidate reflections, achieving sub-pixel Bragg peak localization with lower latency. PeakNet \cite{PeakNet_Peck2025} extends this to a full-image detection framework based on a transformer-style backbone, demonstrating improved recall for weak reflections. Mendez et al. \cite{Mendez2024_DeepResidualNets} train deep residual networks on synthetic nanoBragg data to directly predict crystal hit status and peak positions, showing that simulation-to-real transfer is viable for crystallography. These methods advance detection accuracy; however, they all require GPU inference for frame-rate operation and their weights are not naturally expressible as fixed filter banks, making on-detector ASIC integration impractical with present technology.

\paragraph{Wavelet methods in scientific imaging.}
The DWT was introduced as a tool for multiresolution signal analysis by Mallat \cite{Mallat1989}, and it underlies established image-compression standards (JPEG 2000 uses a biorthogonal CDF 9/7 wavelet). In astronomy and electron microscopy, wavelet-domain thresholding for background separation is well established: the \textit{\`a trous} algorithm decomposes an image into isotropic wavelet planes and suppresses extended smooth emission while preserving compact sources \cite{Starck1994}. To our knowledge, no prior work applies a 2D DWT decomposition pipeline to X-ray serial crystallography for the purpose of real-time, hardware-mappable background segmentation.

\paragraph{On-detector and near-detector processing.}
The need for data reduction at the sensor level is increasingly recognized across X-ray and electron-microscopy detector communities. Rasheedi et al.\ \cite{rasheedi202528} demonstrate a 28-nm multiply-accumulate ASIC that performs on-chip compression at MHz frame rates, using fixed linear transforms expressible as integer-valued filter banks, the same class of operations as a Haar DWT. SparkPix-S \cite{rota2024sparkpix, markovic2023sparkpix} implements pixel-level thresholding directly in the ASIC front-end, reducing output bandwidth by suppressing below-threshold pixels. The present work occupies a complementary niche: rather than per-pixel thresholding or generic compression, we target a structured segmentation that identifies background versus diffraction signal regions, producing subbands that are directly usable by downstream crystallographic pipelines. The DWT analysis filter coefficients are fixed constants (no training), making the compute kernel equivalent to a conv2D layer with known weights, which maps efficiently to both FPGA DSP slices and future readout ASICs. The change in weights and bias enables the use of different wavelet families but again no training is needed. 

\section{Methods}
\label{sec:methods}

This work combines two complementary methodologies to address on-detector data reduction for serial crystallography at MHz repetition rates. First, section~\ref{sec:dwt} introduces a 2D discrete wavelet transform segmentation strategy that exploits the natural separation of smooth background scatter and sharp Bragg peaks in the wavelet domain to isolate diffraction signal directly at the detector level. Second, section~\ref{sec:crystallographic_pipeline} describes the established crystallographic analysis pipeline used to evaluate the fidelity of the segmented data, from peak detection and indexing through post-refinement merging and internal consistency assessment. Together, these two components define a complete evaluation framework: the DWT provides the data reduction front-end whose output quality is validated end-to-end by the crystallographic pipeline against unprocessed reference data.

\subsection{Discrete Wavelet Transform}
\label{sec:dwt}
Wavelets represent signals using localized oscillatory functions at multiple scales and positions. It is considered one of the foundation theories for Multiresolution Analysis (MRA) (unlike the Fourier transform, which uses global sinusoids) where wavelets capture both frequency and spatial/temporal localization. The Wavelet signal decomposition uses a scaling filter and a wavelet filter, the first gives a low resolution version of the input signal while the second computes the differences.  A mother wavelet $ \psi(t) $ is dilated and translated to form a family 
$ \psi_{a,b}(t) = \frac{1}{\sqrt{a}} \psi\left(\frac{t - b}{a}\right), \text{where } a > 0 $
controls the scale and b controls the position. 

Multiresolution analysis formalizes wavelets via nested subspaces $V_j$ that approximate the signal at successively coarser resolutions. The scaling function $\Phi$ (low-pass) spans $V_j$; the wavelet $\psi$ spans the detail space $W_j$ so that $V_{j+1} = V_j \oplus W_j$. The DWT discretizes scales to dyadic values $(a = 2^j)$ and positions to integer steps, yielding a critically sampled, invertible representation. Its implementation uses a two-channel filter bank:
\begin{itemize}
    \setlength\itemsep{0em}
    \item Analysis: convolve with low-pass $h[n]$ (scaling) and high-pass $g[n]$ (wavelet), then downsample by 2 to produce approximation coefficients $a_j$ and detail coefficients $d_j$.
    \item Synthesis: upsample $a_j$ and $d_j$ by 2, convolve with synthesis filters (often related to $h$ and $g$), and sum to reconstruct.
\end{itemize}

DWT can be extended to multiple dimensions; in this work, two-dimensional analysis (DWT2D) is of great interest for the application in imaging detectors. For separable kernel filters the implementation of DWT2D can be done by applying the 1D DWT filters along rows and then columns. For non-separable kernel filters a 2D convolution can be used. One decomposition level yields four subbands:

\begin{itemize}
    \setlength\itemsep{0em}
    \item LL: low-pass in both directions (coarse approximation).
    \item LH: low horizontal resolution, high vertical resolution (vertical edges).
    \item HL: high horizontal, low vertical (horizontal edges).
    \item HH: high-pass in both directions (diagonal details).
\end{itemize}

Multi-level decomposition repeats the transform on the LL subband, producing a pyramid: $LL_j$ plus detail subbands $LH_j$, $HL_j$, $HH_j$ for $j = 1..J$. Here $j$ denotes the decomposition level, corresponding to dyadic scale $a = 2^j$, and $J$ is the total decomposition depth; all results in this work are reported as a function of $J$.

\subsubsection{Serial Crystallography DWT segmentation}
\label{sec:dwt_segmentation}

Wavelet theory provides a set of rules that can be used to define which functions can be considered mother wavelet and scaling function pairs and how they can be used to change the signal being decomposed. Some of the key characteristics of scaling and wavelet transforms that support the segmentation strategy proposed in this work are briefly described in these paragraphs: 

\textit{Localization}: Scaling functions capture smooth, coarse structure which are the background signal in the images under study; Wavelets are localized in time/space and frequency, enabling sharp, position-dependent analysis, and wavelets capture local transitions and edges which is needed to identify where diffraction peaks are located in an image.   

\textit{Multiresolution}: Dyadic scales tile the time-frequency plane so that coarse scales have broader spatial support but narrower frequency bands; fine scales have sharper spatial localization but broader frequency support which means we need to look at a range of scales to capture diffraction peaks of different sizes; 

\textit{Compact support and regularity}: Many practical wavelets (e.g., Haar, Daubechies, biorthogonal spline) have finite support, which controls localization and computational cost. The compact support enables parallel and distributed computation of the filters (i.e. per ASIC filtering as we will describe in the hardware implementation). Regularity (smoothness) influences how well smooth features are represented and compression performance. In this work, the Haar wavelet (db1) is used: it has one vanishing moment, the shortest possible filter support ($h = [1,\,1]/\sqrt{2}$, $g = [1,\,-1]/\sqrt{2}$), and integer-valued filter coefficients with a floating point normalization, if the operations are kept separate, making it well suited for hardware implementation with minimal DSP resources.

\begin{figure}[htp!]
  \centering
  \centerline{\includegraphics[width=12cm]{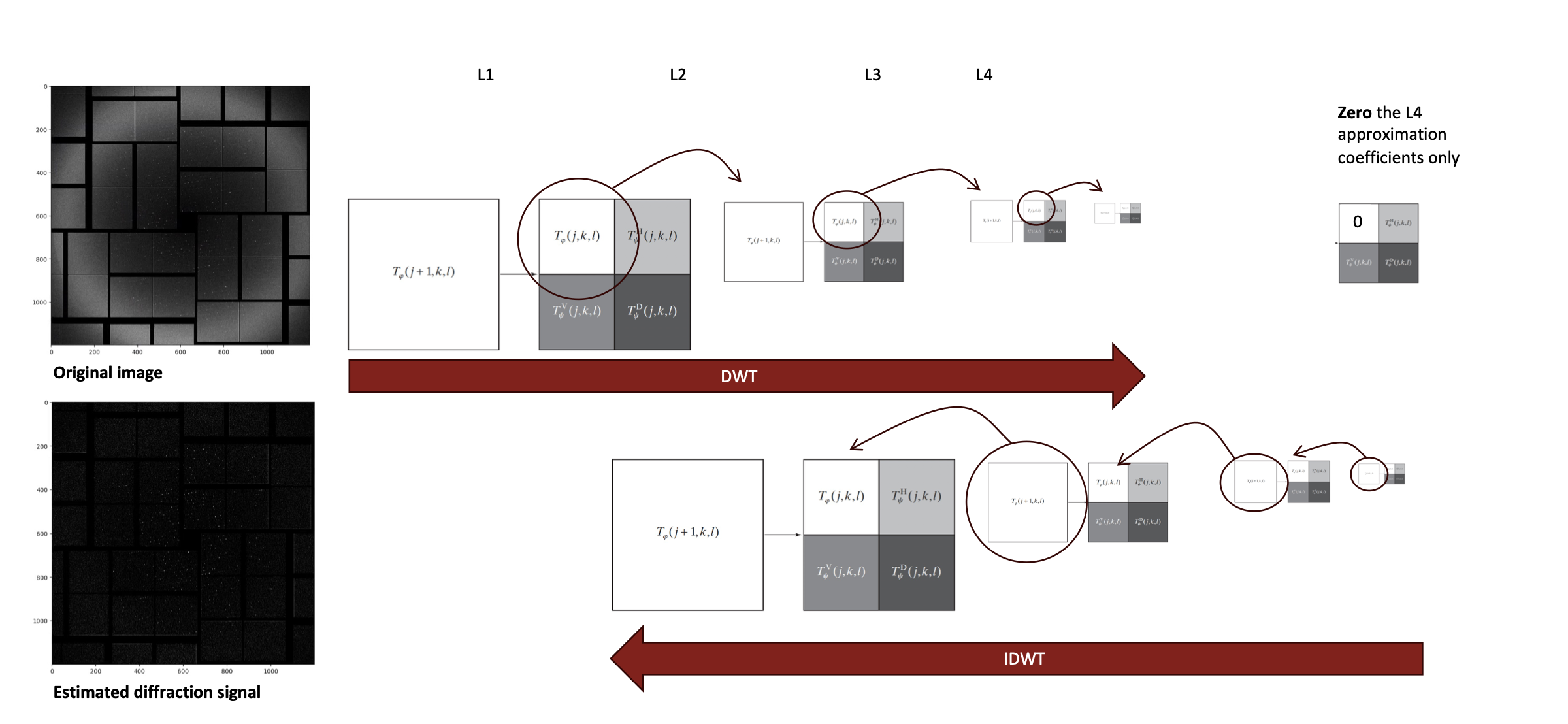}}
  \caption{Diffraction signal segmentation based on DWT2D.}
  \label{fig:dwtsignaldecomposition}
\end{figure}

\textit{Vanishing moments}: A wavelet with M vanishing moments $\int t^k \psi(t) \, dt = 0, \quad \text{for } k = 0, \ldots, M-1$ annihilates polynomials up to degree $M-1$, pushing smooth content into the approximation (scaling) coefficients and concentrating singularities (edges) in detail coefficients. This characteristic essentially represents the core concept of our segmentation strategy, where smooth content is removed from the segmented image before the reconstruction is performed, producing our estimation of the diffraction pattern. Fig.\ \ref{fig:dwtsignaldecomposition} shows the diagram where on the top left side an original image is shown with both background and diffraction peaks. The original image is decomposed 4 times. The approximation coefficients, as expected, based on the wavelet properties contain the information of the background (smooth low frequency components). Once these coefficients are removed the information about the background goes away with them and the reconstructed image contains part of the diffraction signals. For the reconstruction to contain all the diffraction information content, a decomposition using more scales than what is shown in this example is needed.

\begin{figure}[htp!]
  \centering
  \centerline{\includegraphics[width=12cm]{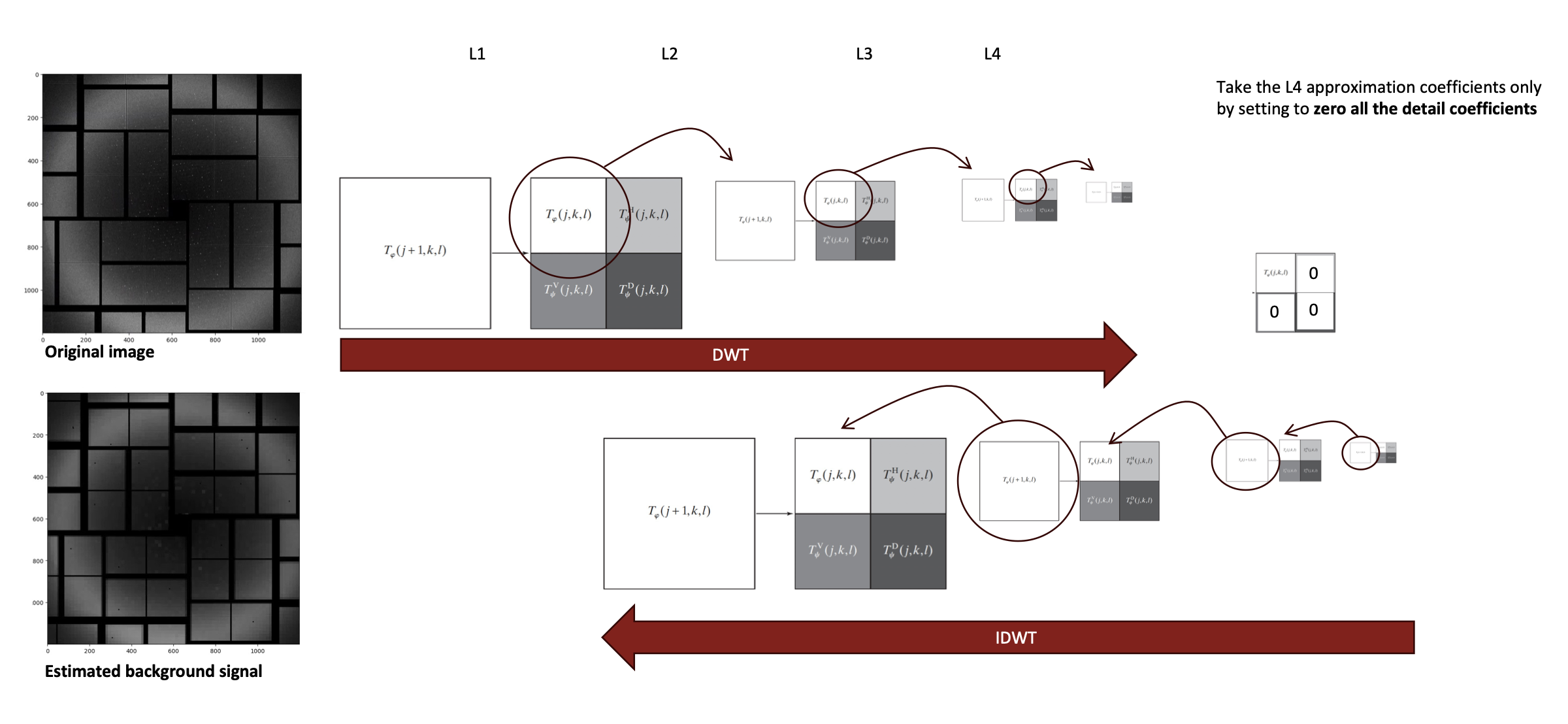}}
  \caption{Background signal segmentation based on DWT2D.}
  \label{fig:dwtbackgrounddecomposition}
\end{figure}

For the background estimation the strategy is reversed, where the detail features are zeroed out before the reconstruction is performed such that the background estimation is produced as presented in Fig.\ \ref{fig:dwtbackgrounddecomposition}. 

The other key characteristic that supports both segmentation strategies is 
\textit{Orthogonality}: Orthonormal sets yield energy preservation and simple inversion essential for the reconstruction of the estimation images.  

\subsection{Crystallographic Analysis Pipeline}
\label{sec:crystallographic_pipeline}

In X-ray serial crystallography, standard metrics to assess the quality of a collected dataset include $CC_{1/2}$, $CC^{*}$ and $R_{\mathrm split}$ \cite{White2012,Karplus2012}. For our study, we compute these metrics after processing diffraction snapshots through a standard serial crystallography pipeline, which we outline in this section. Established serial crystallography software packages such as CrystFEL \cite{White2012} provide the implementation of each step, with tunable parameterization to account for the detector and beam conditions.

\textit{Peak finding and indexing}: Peak detection consists in identifying candidate Bragg reflections using the \texttt{find\_peaks\_v3\_r3} algorithm implemented in the psalgos library \cite{psalgos}. Following peak detection on diffraction image $j$, the experimental geometry maps the detector coordinates of the $i^{\mathrm{th}}$ detected Bragg peak to the scattering vector $\mathbf q_i^{(j)}$ on the Ewald sphere. Assuming that the reciprocal-space basis matrix $B\in\mathbb R^{3\times3}$, determined \textit{a priori} from the crystal unit cell, is known, the indexing algorithm estimates the crystal orientation $U_j\in SO(3)$ and assigns a Miller index $\mathbf h_i^{(j)}\in\mathbb Z^3$ to each observed scattering vector according to
\[
\mathbf q_i^{(j)} \approx U_jB\mathbf h_i^{(j)},
\]
where $B$ maps Miller indices to reciprocal-space vectors in the crystal reference frame, and $U_j$ rotates these vectors into the laboratory frame \cite{BusingLevy1967}. The \texttt{indexamajig} program from the CrystFEL suite \cite{White2012} estimates the crystal orientation, which is defined only up to the point-group symmetry of the crystal \cite{BrehmDiederichs2014} by employing sequential indexing attempts with \texttt{XDS} \cite{Kabsch2010} and \texttt{MOSFLM} \cite{Powell1999}. 


\textit{Integration, scaling and merging}: For successfully indexed patterns, \texttt{indexamajig} predicts the positions of indexed Bragg reflections and integrates the corresponding reflection intensities. Because each still diffraction pattern records only a thin slice of reciprocal space, the integrated reflection intensity  $\{I_i^{(j)}\}$ of the $i^{\mathrm{th}}$ indexed reflection on diffraction image $j$ obtained by \texttt{indexamajig} represents partial measurements of the underlying Bragg reflection intensities. The \texttt{partialator} program from the CrystFEL suite \cite{White2016} refines per-crystal scale factors and reflection partialities across all indexed diffraction images. It then combines observations whose assigned Miller indices $\mathbf h_i^{(j)}$ coincide. Specifically, observations satisfying $\mathbf h_i^{(j)}=(h,k,l)^\top$ correspond to the same crystallographic reflection and contribute to a single merged intensity $\hat I_{hkl}$, which is proportional to the squared structure-factor amplitude $|F_{hkl}|^2$.

\textit{Data quality assessment}: To assess data quality, \texttt{partialator} splits the diffraction measurements into two disjoint half-datasets and merges each half independently, producing merged reflection intensities estimates $\{\hat I_{hkl}^{(1)}\}$ and $\{\hat I_{hkl}^{(2)}\}$ with respective means $\overline{\hat I}^{(1)}$ and $\overline{\hat I}^{(2)}$. The \texttt{compare\_hkl} program from CrystFEL compares these independently merged reflection intensities by computing the split-dataset residual,

\begin{equation*}
R_{\mathrm{split}} =
\frac{1}{\sqrt{2}}
\frac{\sum_{hkl} \left|\hat I_{hkl}^{(1)}-\hat I_{hkl}^{(2)}\right|}
{\tfrac{1}{2}\sum_{hkl}\left(\hat I_{hkl}^{(1)}+\hat I_{hkl}^{(2)}\right)},
\end{equation*}

as defined in \cite{White2012}. The half-dataset correlation coefficient is computed as

\begin{equation*}
CC_{1/2} =
\frac{\sum_{hkl}
\left(\hat I_{hkl}^{(1)}-\overline{\hat I}^{(1)}\right)
\left(\hat I_{hkl}^{(2)}-\overline{\hat I}^{(2)}\right)}
{\sqrt{
\sum_{hkl}\left(\hat I_{hkl}^{(1)}-\overline{\hat I}^{(1)}\right)^2
\sum_{hkl}\left(\hat I_{hkl}^{(2)}-\overline{\hat I}^{(2)}\right)^2}},
\end{equation*}
and converted to the correlation coefficient of the fully merged dataset as introduced in \cite{Karplus2012}

\begin{equation*}
CC^{*}=\sqrt{\frac{2CC_{1/2}}{1+CC_{1/2}}}.
\end{equation*}

These metrics quantify the internal consistency of the merged dataset and guide the selection of the resolution cutoff.
\section{Results}
\label{sec:results}

This section evaluates the DWT segmentation strategy from multiple complementary perspectives. Section~\ref{sec:results_dwt_segmentation} presents qualitative segmentation results on both real and simulated data and introduces the pixel-level mismatch analysis used to characterize reconstruction fidelity. Section~\ref{sec:results_pf8} benchmarks the DWT against the established peakfinder8 algorithm using peak-level precision and recall on simulated data. Section~\ref{sec:results_wavelets} compares 12~wavelet families to justify the choice of the Haar filter. Section~\ref{sec:results_noise} examines noise robustness under increasing levels of additive Gaussian noise, exposing the current limitations of the pipeline. Finally, section~\ref{sec:results_cc} assesses the impact of decomposition depth on downstream crystallographic data quality using CC* and $R_\mathrm{split}$ metrics, confirming that the segmentation preserves the information content required for structure determination.

\subsection{DWT Segmentation Results}
\label{sec:results_dwt_segmentation}
\begin{figure}[htp!]
    \centering
    \subfloat[]{
        \includegraphics[width=0.3\textwidth]{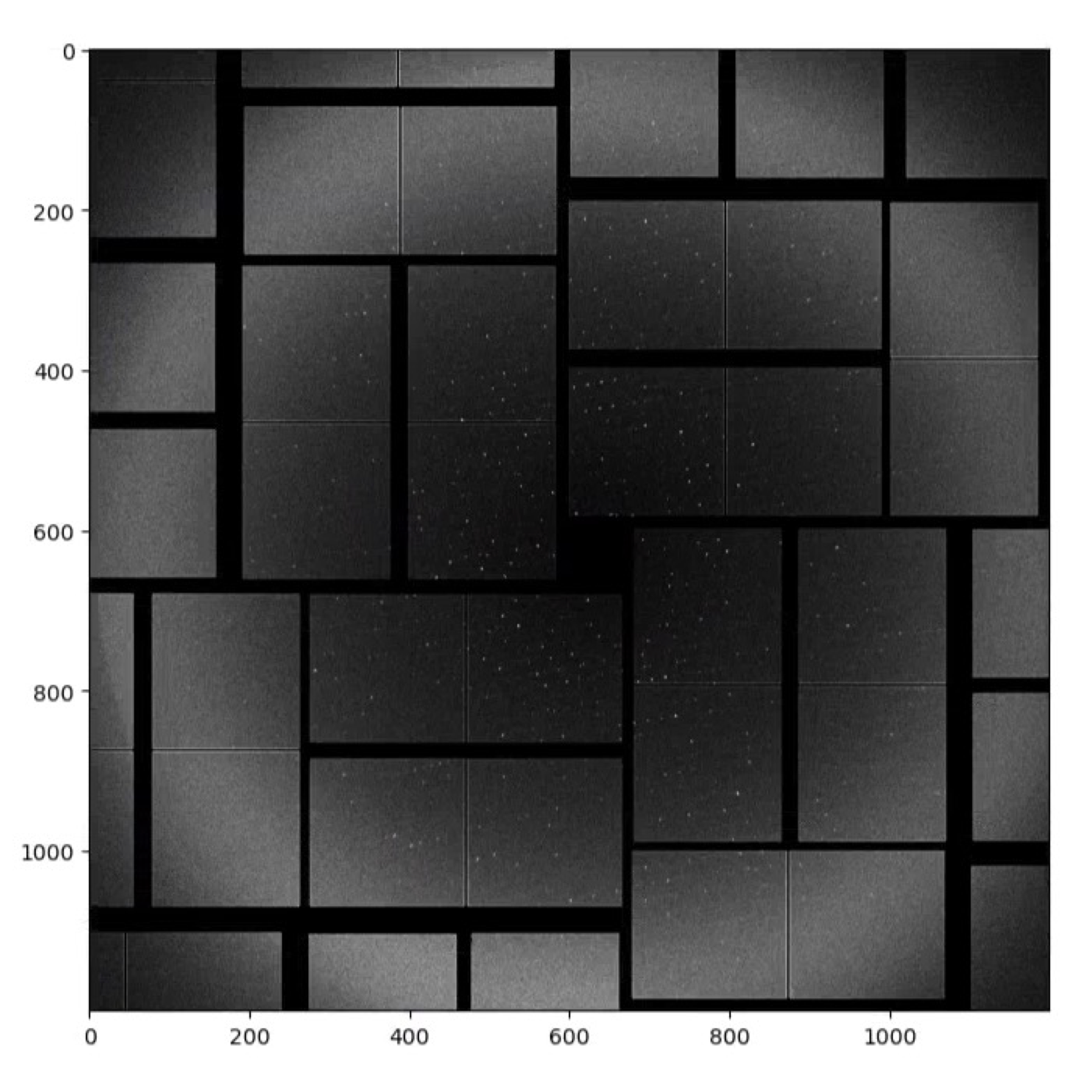}}
    \subfloat[]{
        \includegraphics[width=0.3\textwidth]{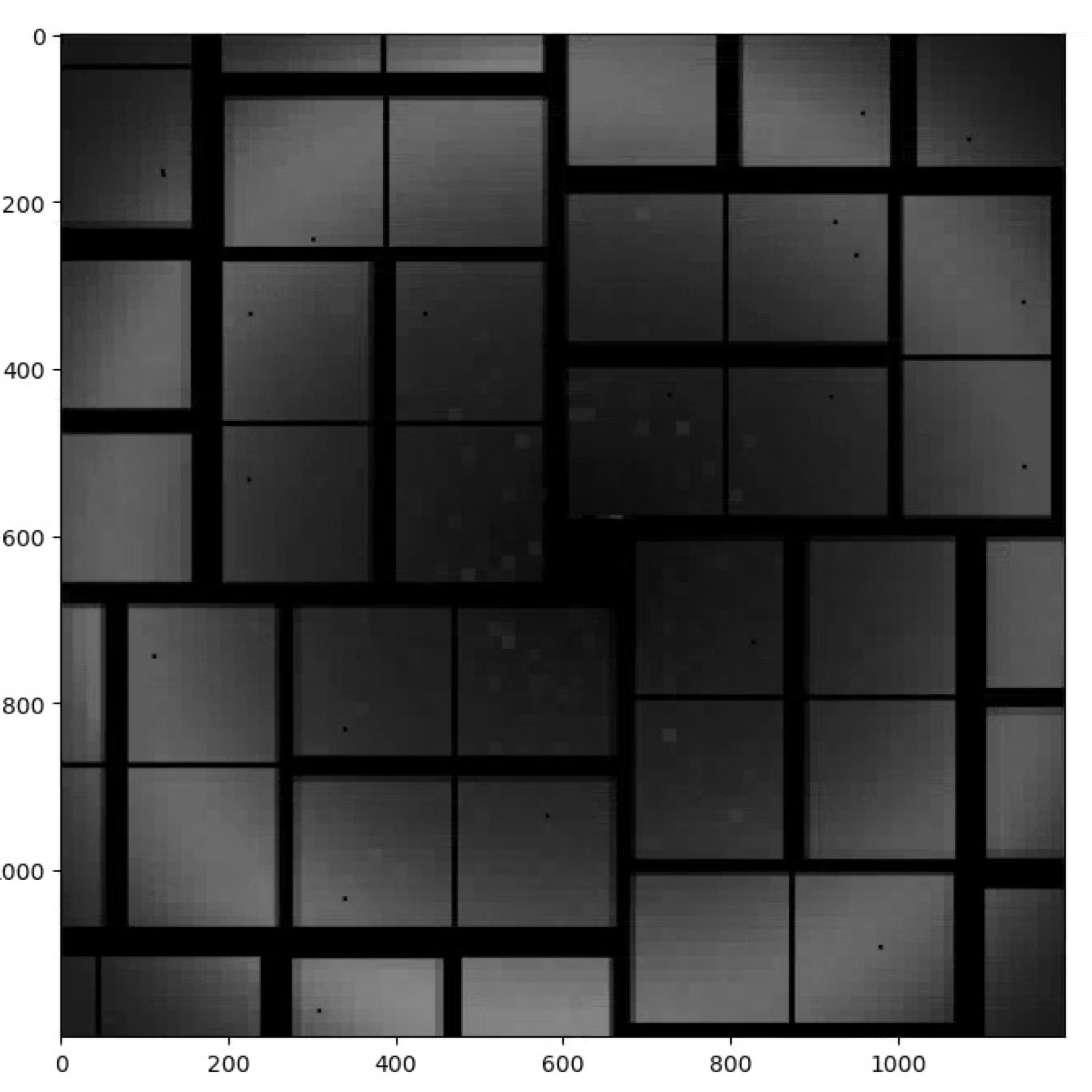}}
    \subfloat[]{
        \includegraphics[width=0.3\textwidth]{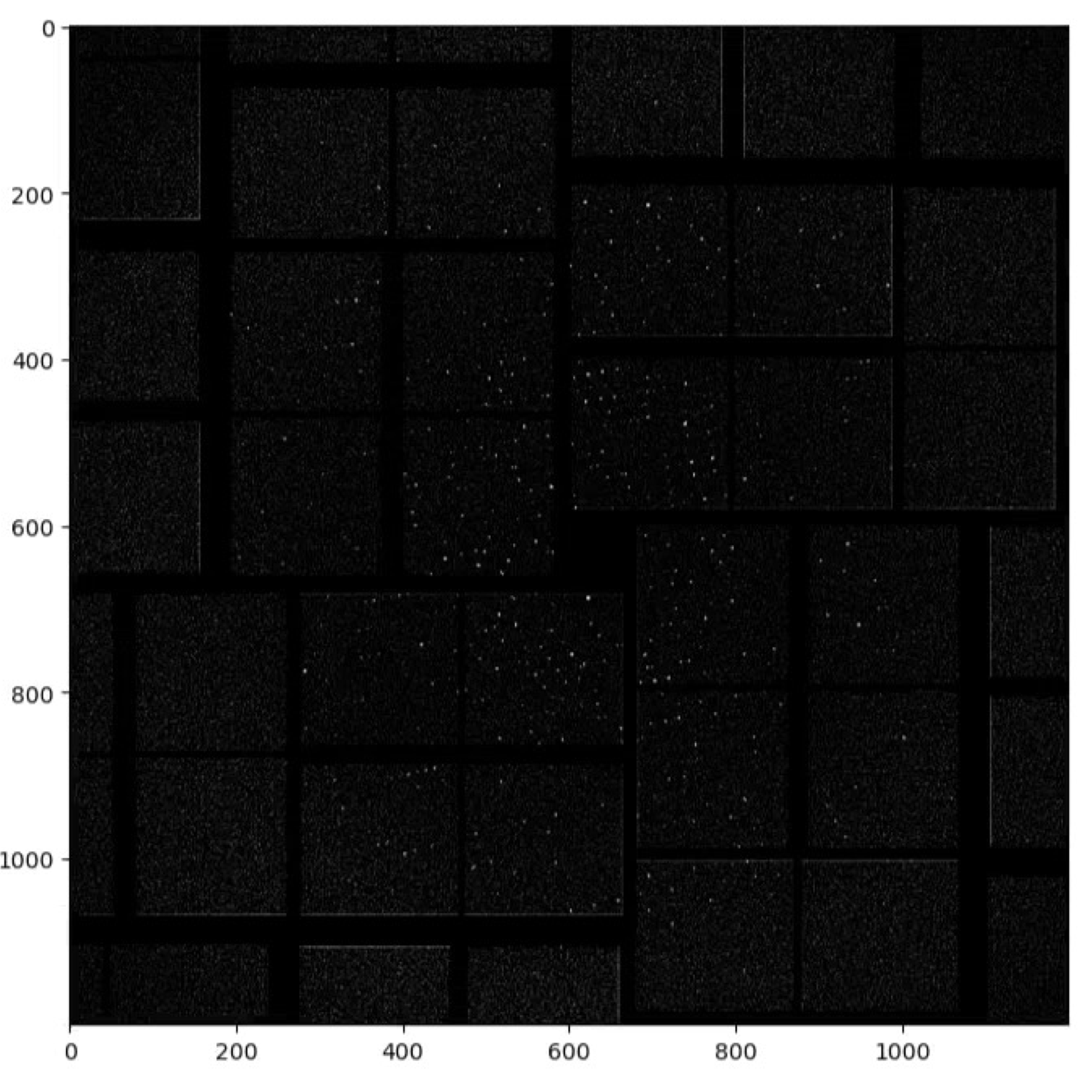}}
    \caption{Segmentation example using a real image from the CSPAD detector taken at LCLS \cite{ke2018convolutional} (a) Original image. (b) Background signal estimate. (c) Diffraction signal estimate.}
    \label{fig:segmenationCSPADexample}
\end{figure}

This subsection presents the results obtained by the DWT segmentation as proposed, using the Haar wavelet (db1) for all experiments. The initial result is for a real image taken at LCLS for a serial crystallography experiment. In this experiment the images were calibrated before they were used for classification where an image with clear diffraction peaks was classified as HIT, an image with only background information and no diffraction signal was classified as MISS and images where there were less than 10 diffraction spots were classified as MAYBE. Fig. \ref{fig:segmenationCSPADexample} shows the original image, the background estimation and the diffraction signal estimation for a HIT image. This result is a qualitative example that shows that the smooth background can be segmented from the diffraction peaks as described in section \ref{sec:dwt_segmentation}.

\begin{figure}[htp!]
    \centering
    \subfloat[]{
        \includegraphics[width=0.307\textwidth]{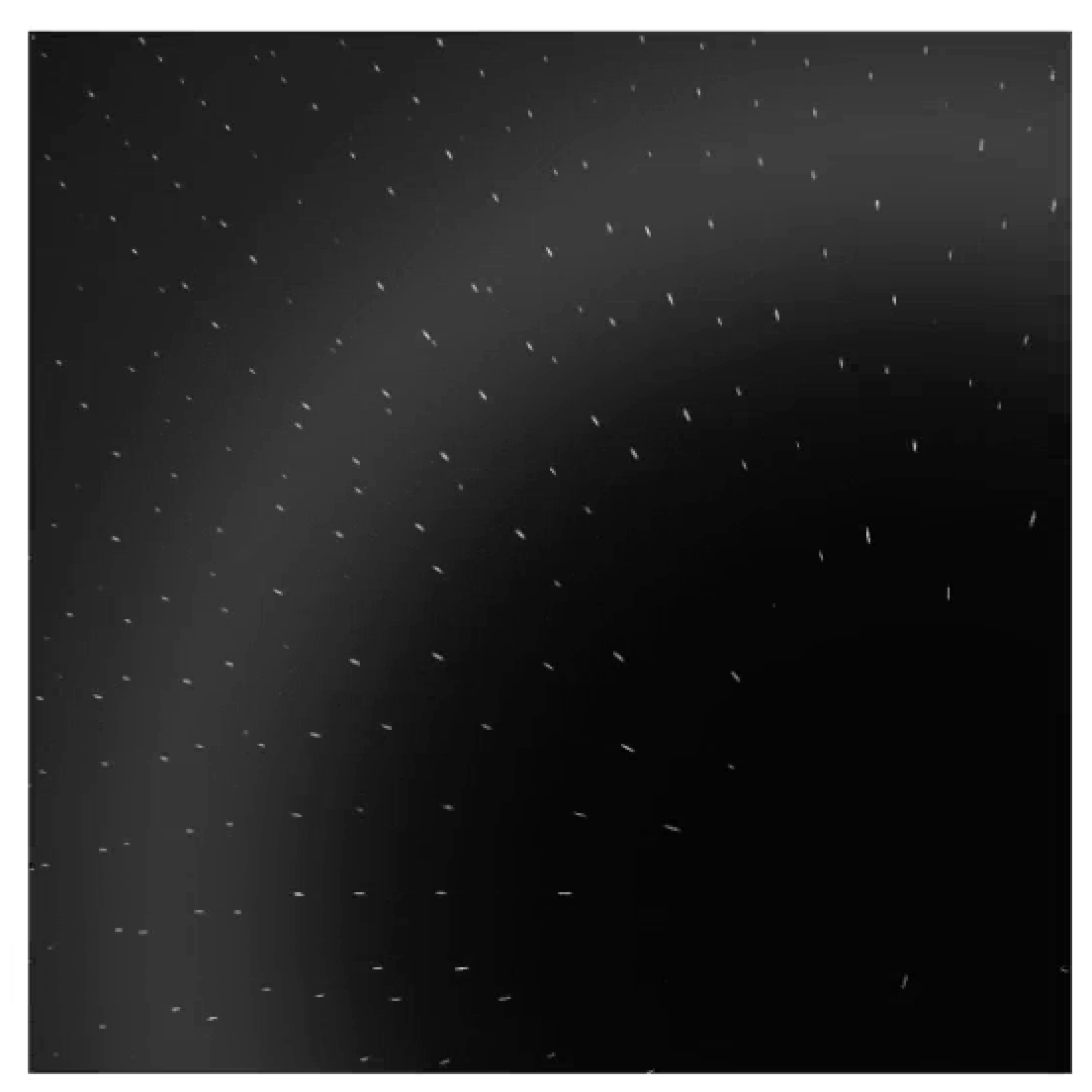}}
    \subfloat[]{
        \includegraphics[width=0.3\textwidth]{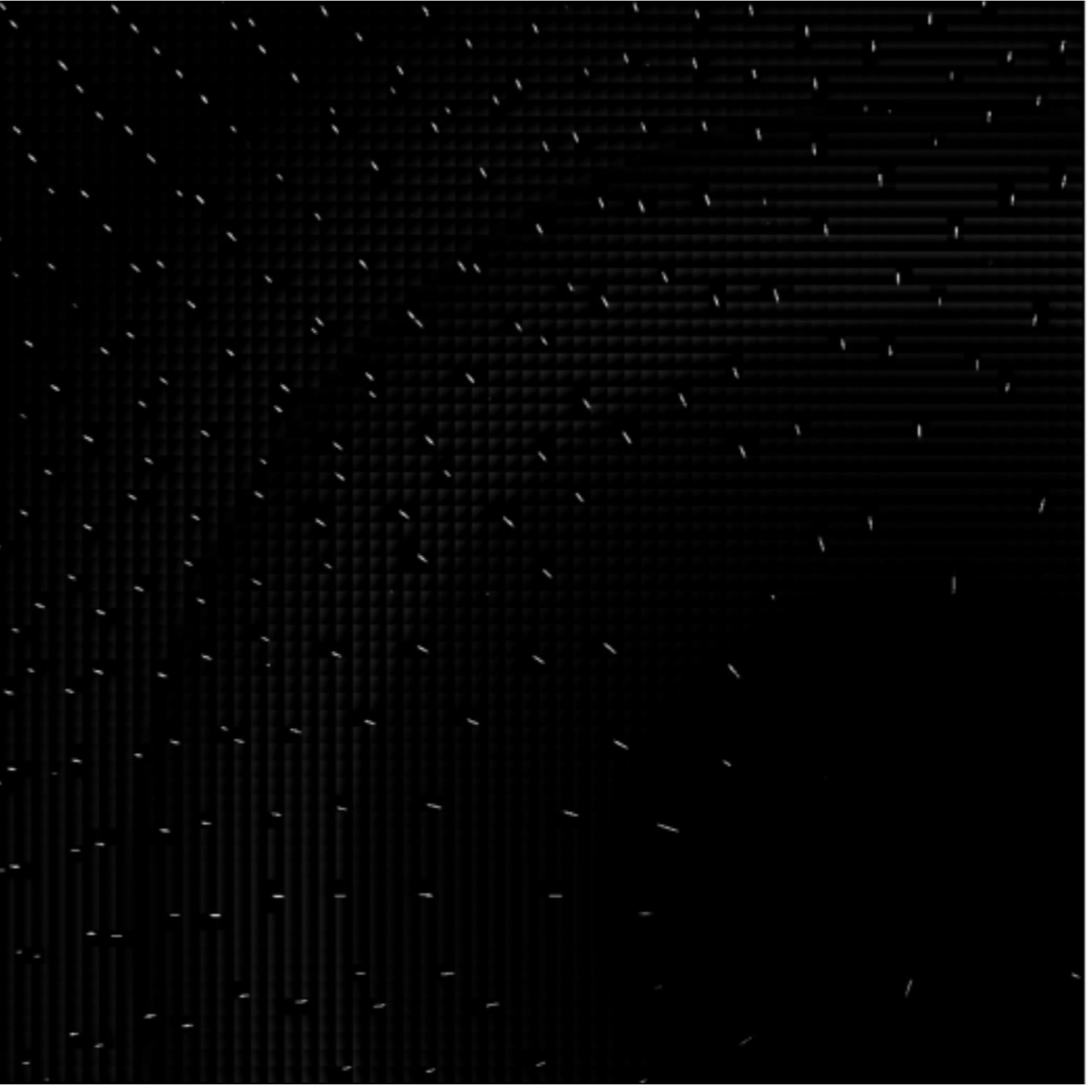}}
    \subfloat[]{
        \includegraphics[width=0.3\textwidth]{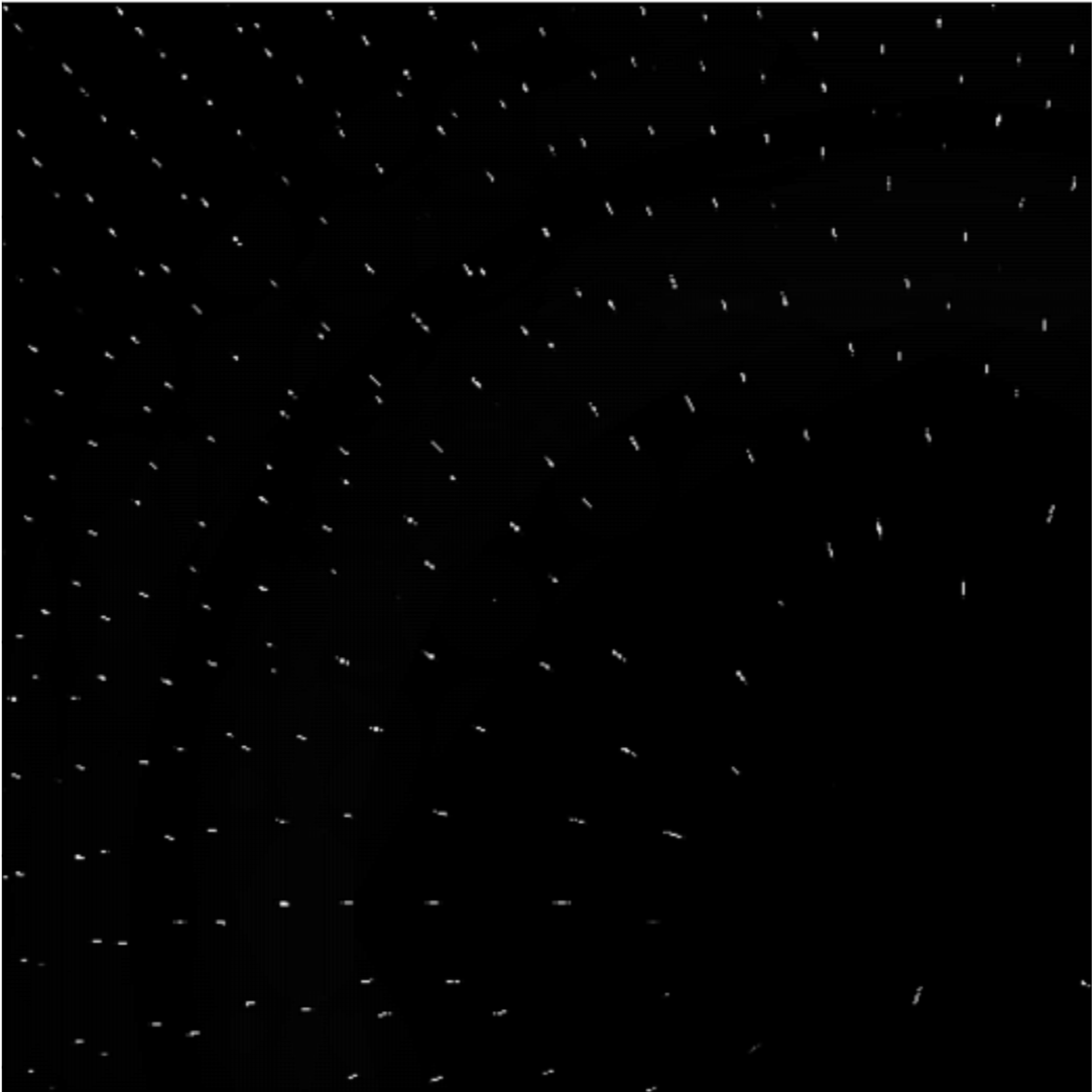}}
    \caption{Segmentation example using simulated data from \texttt{nanoBragg}. (a) Original image with background and Bragg peaks. (b) Estimated diffraction signal using 4 levels of DWT decomposition (detail subbands only). (c) Full reconstruction showing the combined background and diffraction estimate.}
    \label{fig:segmenationSimExample}
\end{figure}

To validate and benchmark the proposed DWT-based hit-finding and Bragg-peak segmentation under controlled yet experiment-relevant conditions, we generate a physics-realistic synthetic diffraction dataset using a simulation framework built on \texttt{nanoBragg} forward model within the \texttt{simtbx} toolkit (as part of the Computational Crystallography Toolbox ecosystem) 
coupled to \texttt{dxtbx} for the detector and experiment-geometry representation \cite{Young2023_Chp7, Parkhurst2014_dxtbx}. 
This approach, as shown in Figure \ref{fig:segmenationSimExample}, enables controlled sweeps over experimental parameters (crystal size/mosaicity, sample to detector distance and beam center, background scattering, and noise) 
and detector type/configuration choices that can be used to synthesize diverse training and test sets. 

To further evaluate the proposed DWT-based segmentation, we selected a set of 10 images and estimated their diffraction patterns. From simulation, we obtained the corresponding background-free diffraction patterns and used these as golden references. Both the measured and reference images were binarized using a single global threshold. A pixel-wise XOR between the two binary images identified the pixels where the two assessments disagreed.

\begin{figure}[htp!]
    \centering
    \subfloat[]{
        \includegraphics[width=0.307\textwidth]{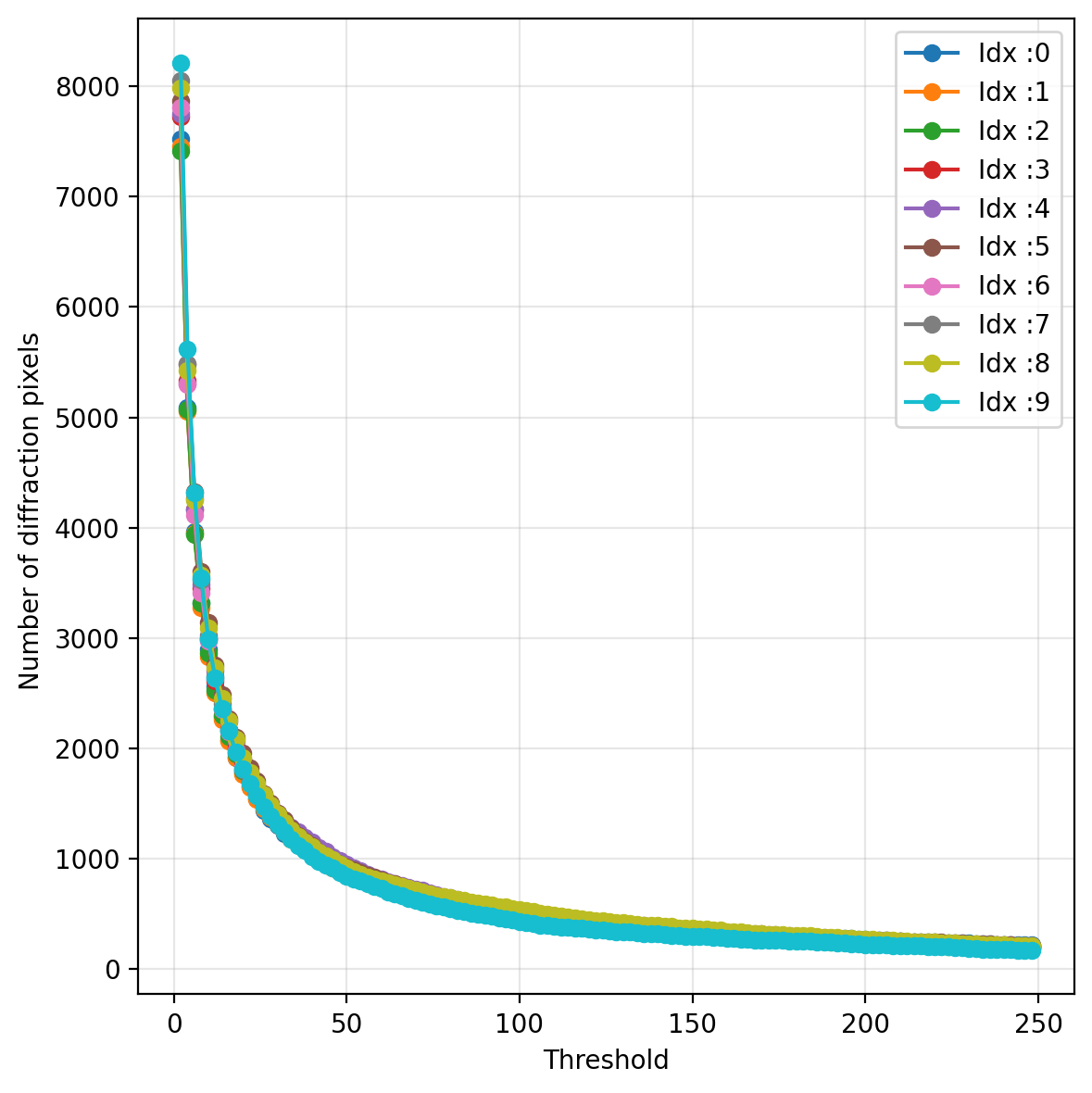}}
    \subfloat[]{
        \includegraphics[width=0.3\textwidth]{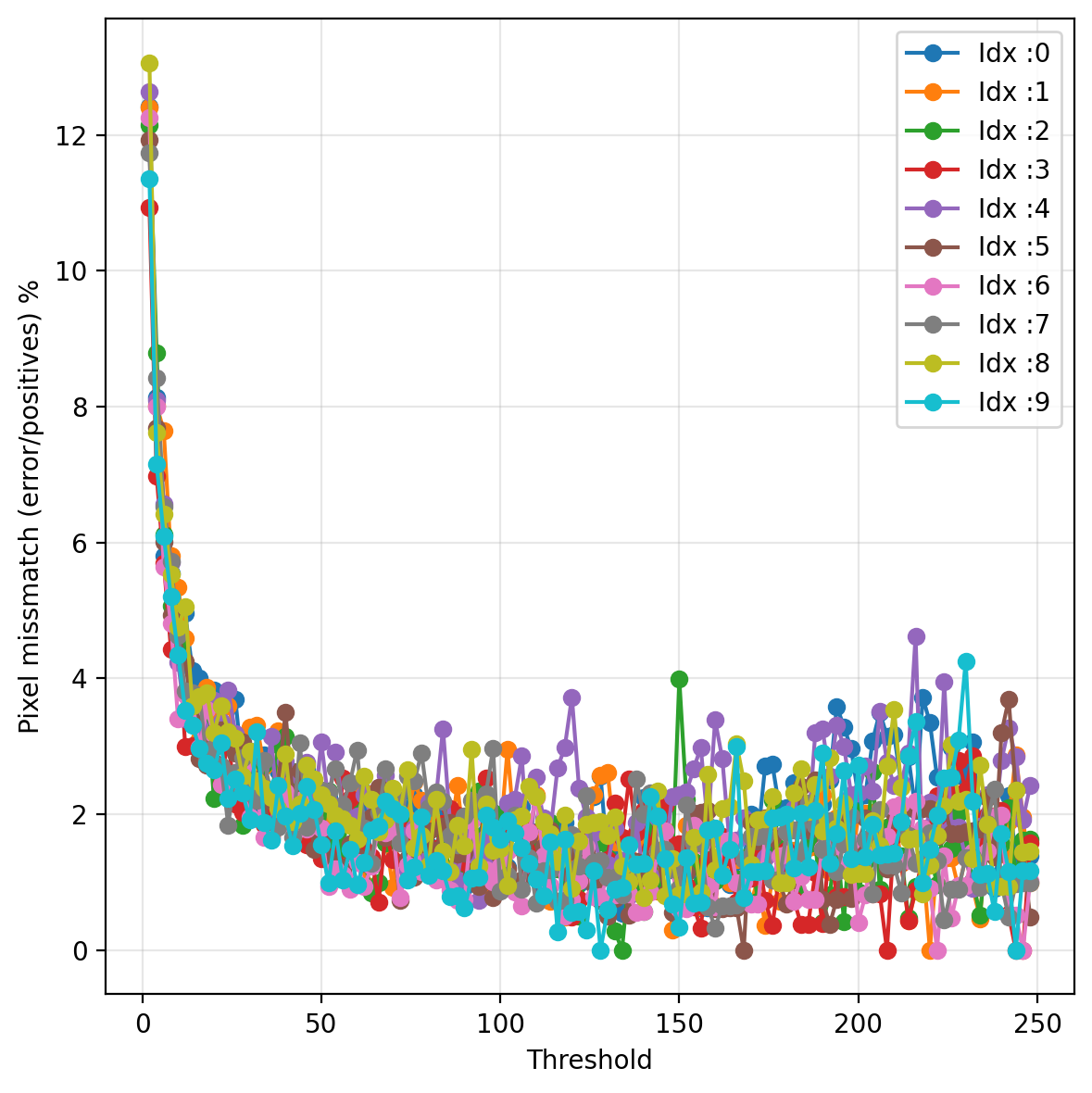}}
    \subfloat[]{
        \includegraphics[width=0.3\textwidth]{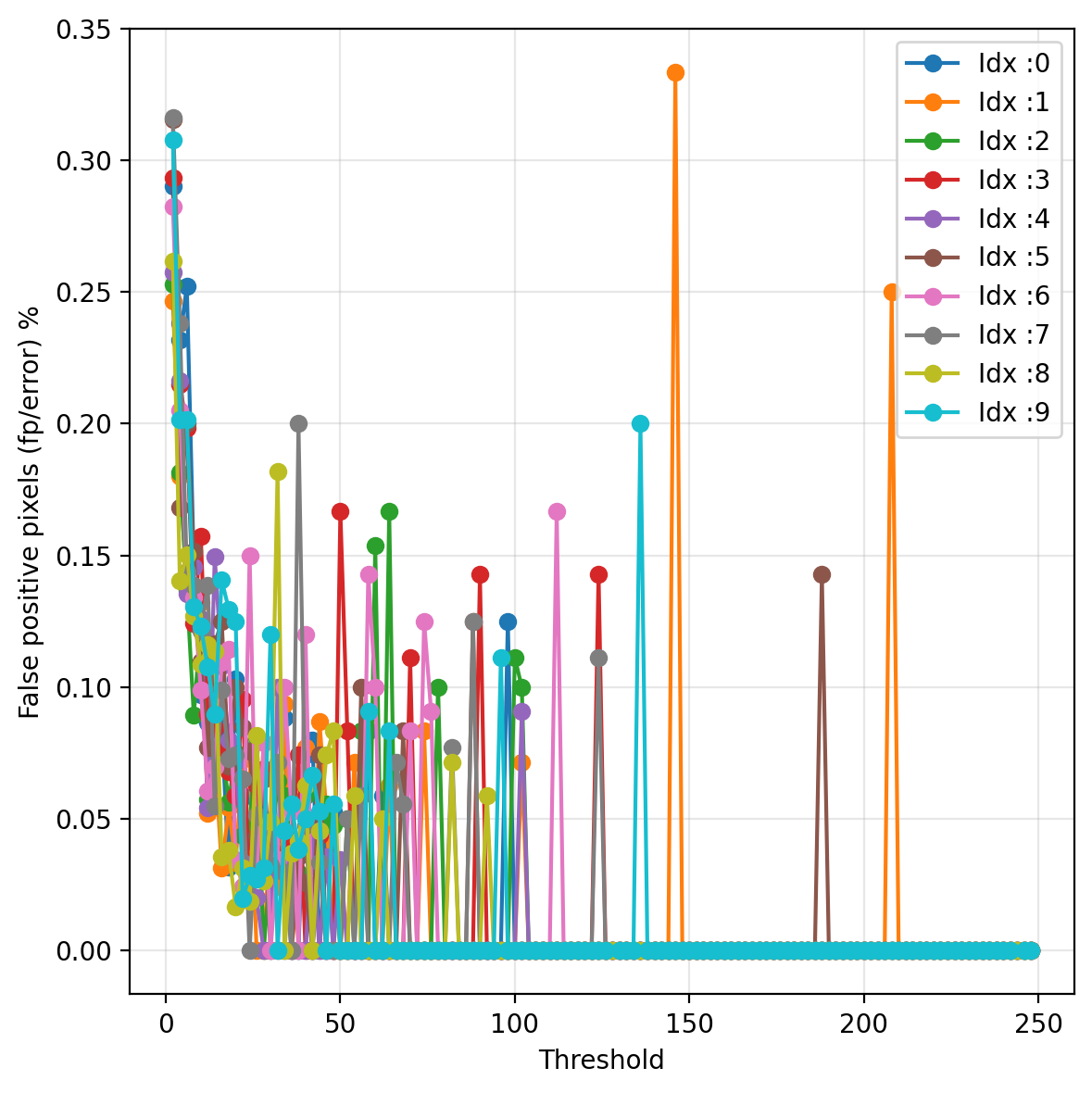}}
    \caption{Mismatch pixel localization based on simulated image (a) Number of diffraction pixel from background free image. (b) Number of pixels that were reported as error (mismatch) when comparing the position of the diffraction pixels of the reconstructed image and the background free image. (c) Percent of false positive identified pixels.}
    \label{fig:quantitavieResultsRoi1}
\end{figure}

For the selected dataset, Figure \ref{fig:quantitavieResultsRoi1} shows: (a) the number of pixels classified as diffraction, (b) the percentage of misclassified pixels and (c) the number of false positive pixels segmented in the image set. In the mismatch plot, the error is initially high, reflecting difficulty in localizing low-intensity peaks. As the threshold suppresses these small values, the error falls below 4\%. However, when the threshold is increased to retain only high-intensity pixels, the error rises again, likely due to limited statistics, as the total number of retained pixels decreases monotonically with increasing threshold.

\subsection{Peak-Level Comparison: DWT2D vs.\ Peakfinder8}
\label{sec:results_pf8}

A final test performed was to compare the DWT-based segmentation against the established peakfinder8 algorithm~\cite{Barty2014}, in it we evaluate both methods on 
100 simulated nanoBragg frames (with a mean/minimum/maximum number of peaks of 1647.8/1543/1801 per golden) image using peak-level precision and recall. Ground-truth peak positions are extracted from the noiseless diffraction images via connected-component labeling; a detected peak is counted as a true positive if it falls within 5~pixels of a ground-truth peak. At this point we consider the DWT a potential in hardware peak pre-selection algorithm and pf8 as a second step to filter the valid peaks. This opens the possibility of zero suppression in real-time in firmware or at the ASIC level but the full evaluation of this strategy is considered a future work.

\begin{figure}[htp!]
    \centering
    \includegraphics[width=0.65\textwidth]{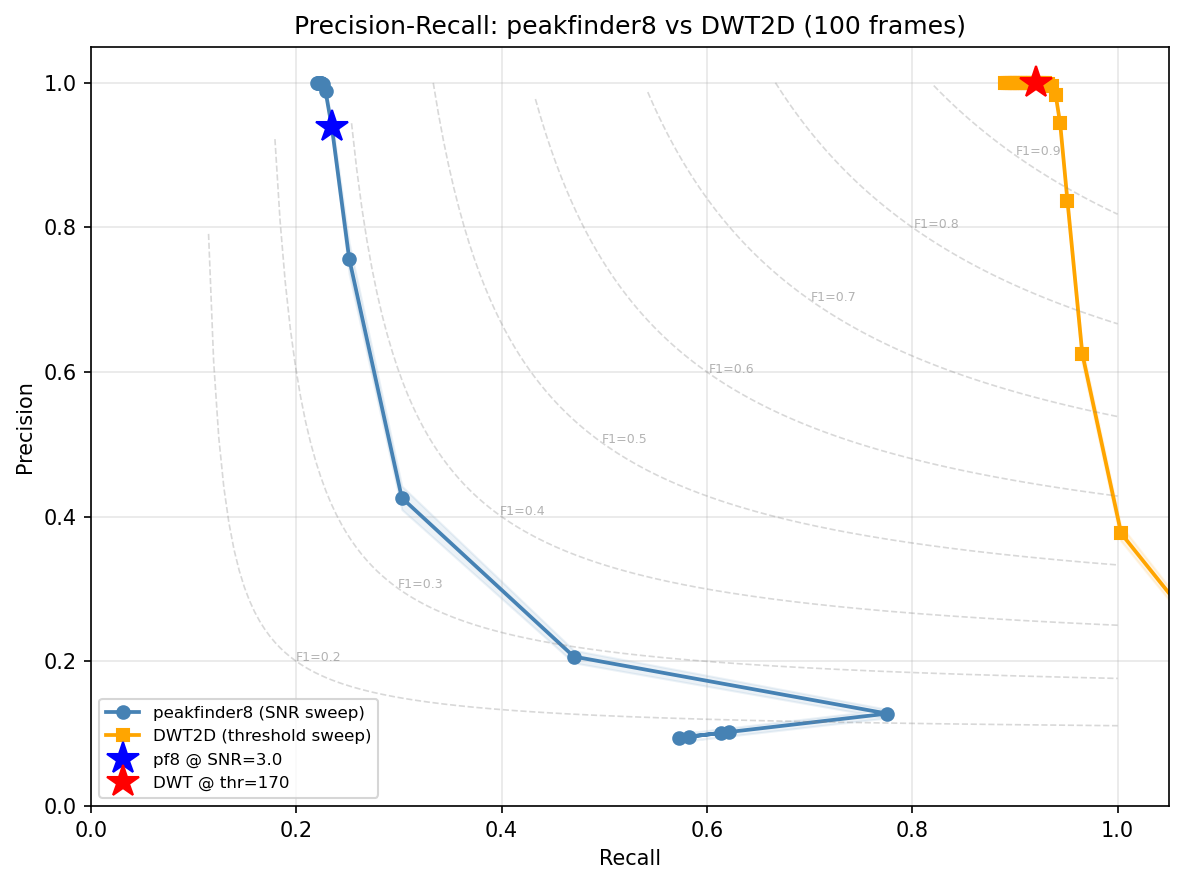}
    \caption{Precision--recall curves for peakfinder8 (SNR sweep) and DWT2D (threshold sweep) evaluated on 100 simulated frames. Star markers indicate the default operating points. F1 iso-contours are shown as dashed grey lines. DWT2D consistently occupies the high-precision, high-recall region of PR space.}
    \label{fig:combined_pr}
\end{figure}

Figure~\ref{fig:combined_pr} shows the precision--recall trade-off for both methods. The DWT2D curve is obtained by sweeping the reconstruction threshold, while the peakfinder8 curve is obtained by sweeping the minimum SNR parameter. At the selected operating points (DWT threshold~$= 170$~photons, pf8 $\mathrm{SNR_{min}} = 3.0$), DWT2D achieves $P \approx 1.00$, $R \approx 0.92$ (F1~$\approx 0.96$), while peakfinder8 achieves $P \approx 0.94$, $R \approx 0.24$ (F1~$\approx 0.37$). The DWT2D method dominates the upper-right region of the PR space across the full parameter sweep, indicating consistently higher recall at comparable or better precision.

\subsection{Wavelet Family Comparison}
\label{sec:results_wavelets}

\begin{figure}[htp!]
    \centering
    \includegraphics[width=0.65\textwidth]{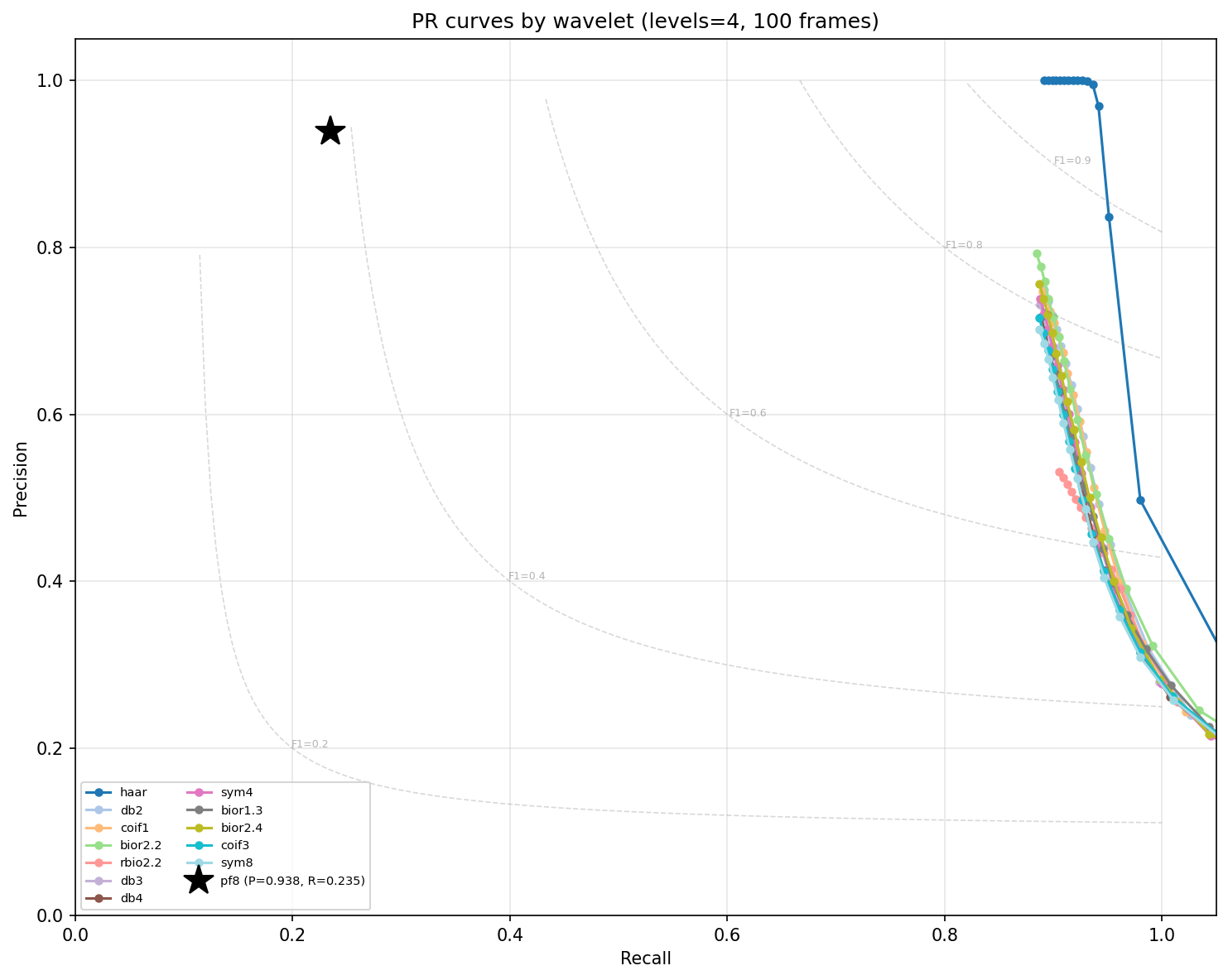}
    \caption{Precision--recall curves for 12 wavelet families at $J = 4$ decomposition levels and 100 frames. The Haar wavelet (blue, solid) is clearly separated from all other wavelets, reaching $F1 > 0.9$. The black star marks the peakfinder8 operating point ($P = 0.94$, $R = 0.24$).}
    \label{fig:pr_wavelets}
\end{figure}

To justify the choice of wavelet filter, we compare 12~wavelet families at fixed decomposition depth ($J = 4$) and threshold (170~photons) across 100 frames. Figure~\ref{fig:pr_wavelets} shows the precision--recall curves obtained by sweeping the DWT reconstruction threshold for each wavelet. The Haar wavelet (db1) clearly dominates, achieving the highest F1~$\approx 0.96$ and uniquely reaching the upper-right corner of PR space. All other wavelets cluster in the $F1 \approx 0.6$--$0.73$ range, with the next-best performers being coif1, bior2.2, and db2. The advantage of Haar is consistent with theoretical expectations: its 2-tap filter provides the shortest support (1~pixel at level~0), maximally preserving the sharp, localized structure of Bragg peaks, while longer filters spread peak energy across more coefficients and reduce the reconstructed peak amplitude.

\subsection{Noise Robustness}
\label{sec:results_noise}

\begin{figure}[htp!]
    \centering
    \includegraphics[width=0.65\textwidth]{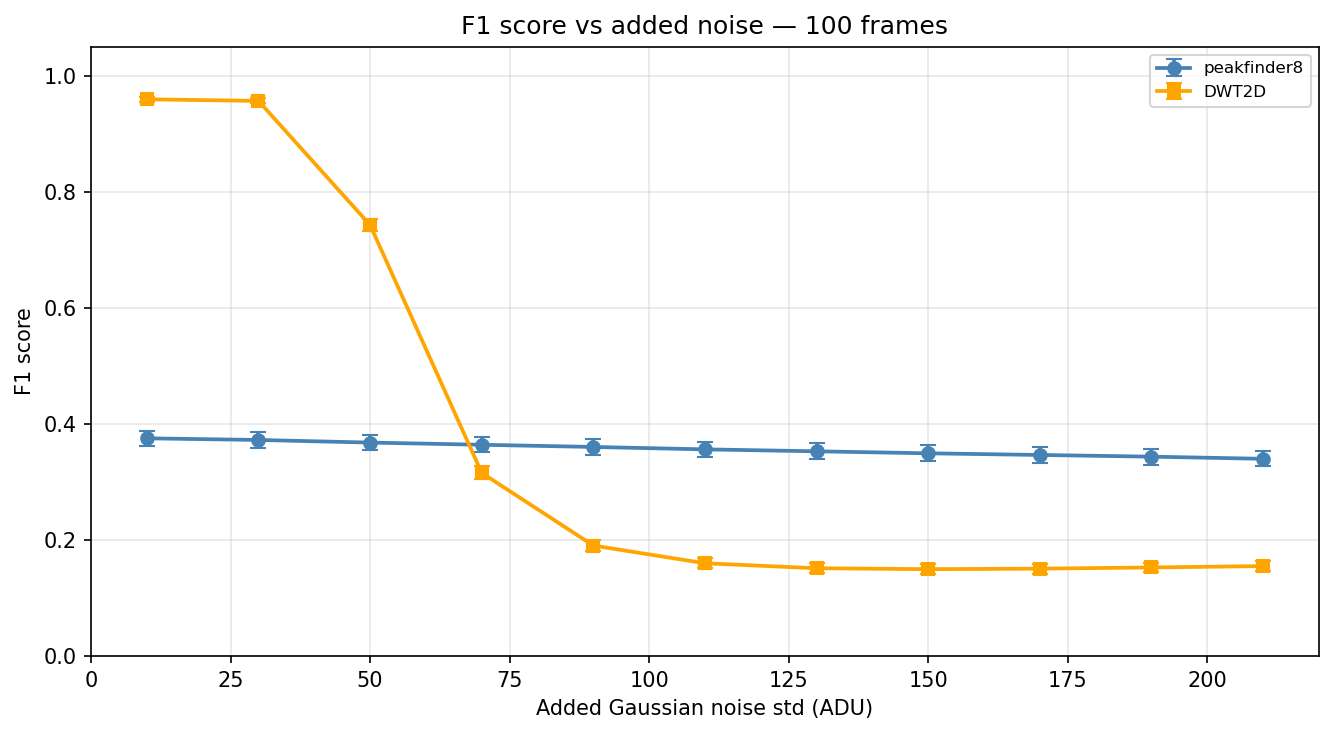}
    \caption{F1 score as a function of added Gaussian noise for DWT2D and peakfinder8 over 100 frames. DWT2D dominates at low noise but degrades sharply beyond $\sim$50~ADU due to noise-induced false positives. Peakfinder8 is noise-robust but operates at a lower F1 baseline due to its inherently lower recall.}
    \label{fig:noise_f1}
\end{figure}

A critical consideration for on-detector deployment is robustness to electronic noise. We evaluate both methods under increasing levels of additive Gaussian noise (zero mean, standard deviation 10--210~ADU, with 1~ADU representing approximately 1~photon) applied to the 100-frame dataset. Figure~\ref{fig:noise_f1} shows the F1 score as a function of added noise standard deviation.

At low noise ($\sigma < 30$~ADU), DWT2D maintains its advantage with $F1 \approx 0.96$. However, as noise increases past $\sim$50~ADU, DWT2D F1 degrades sharply due to a collapse in precision: the simple global threshold on the reconstructed image can no longer distinguish noise-induced artifacts from real peaks. Beyond $\sim$75~ADU, DWT2D F1 falls below peakfinder8. In contrast, peakfinder8 maintains a stable $F1 \approx 0.37$ across all noise levels, reflecting its adaptive radial background and noise estimation. This result highlights a known limitation of the current DWT pipeline, where the LL-zeroing removes background but does not denoise, since all wavelet coefficients are used in the reconstruction. The noise level at which DWT-based segmentation begins to degrade is well above the expected noise floor of the ePixUHR detector (less than 1~photon), so the method remains robust under realistic operating conditions. Should future experiments encounter higher noise, wavelet-domain detail coefficient thresholding (Donoho universal threshold) offers a natural extension to improve robustness.

\subsection{Crystallographic Data Quality as a Function of DWT Decomposition Level}
\label{sec:results_cc}

\begin{figure}[htp!]
    \centering
    \subfloat[CC* vs. resolution]{
        \includegraphics[width=0.48\textwidth]{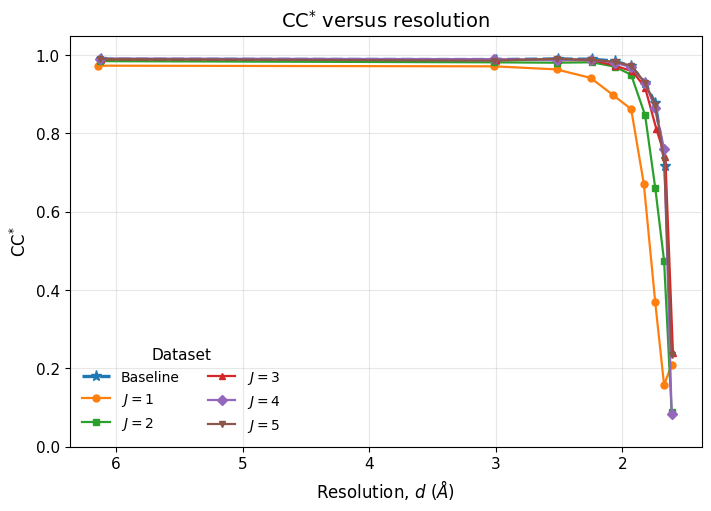}}
    \subfloat[$R_\mathrm{split}$ vs. resolution]{
        \includegraphics[width=0.48\textwidth]{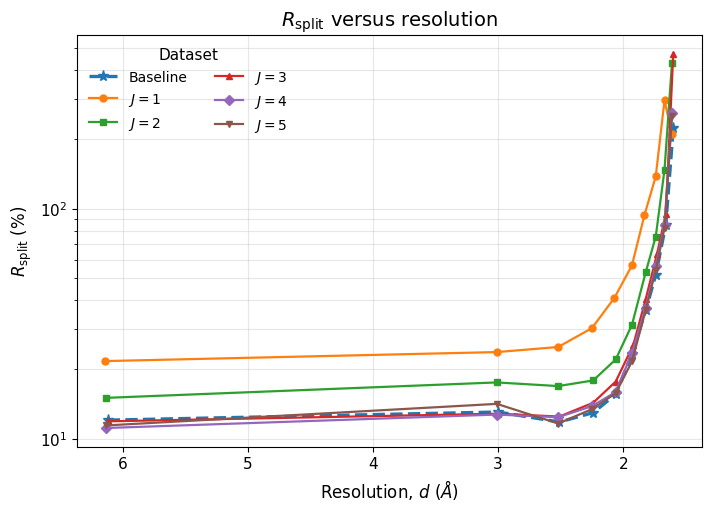}}
    \caption{Crystallographic data quality metrics as a function of resolution for DWT decomposition levels $J = 1$ to $5$ with the unprocessed baseline overlaid. (a) CC* curves: higher is better. (b) $R_\mathrm{split}$ curves: lower is better. Both metrics converge at $J = 4$; levels 4 and 5 match or approach the baseline at mid-resolution. The divergence at the highest-resolution shell (1.6~\AA) reflects low reflection counts in that bin.}
    \label{fig:ccstar_rsplit_all}
\end{figure}

To assess the impact of DWT decomposition depth on downstream crystallographic data quality, we processed a real serial crystallography dataset (ePix10kA detector collecting data at LCLS) using the DWT-segmented images at five decomposition levels ($J = 1$ to $5$) and evaluated the merged structure factor amplitudes using the standard quality metrics CC* and $R_\mathrm{split}$ as a function of resolution. Figure~\ref{fig:ccstar_rsplit_all} shows the resulting curves together with the unprocessed baseline, and Table~\ref{tab:ccstar_rsplit_summary} summarises values at selected resolution shells.

The results reveal a clear monotonic improvement in data quality from $J=1$ to $J=4$. At $J=1$, the decomposition is insufficient to separate background from diffraction signal: CC* degrades sharply beyond 2.5~\AA, reaching 0.16 at 1.7~\AA, and $R_\mathrm{split}$ exceeds 90\% at 1.8~\AA. This is consistent with the theoretical expectation: a single decomposition level removes only the coarsest background component while leaving residual smooth structure in the detail coefficients that contaminates the reconstructed diffraction signal.

Increasing the decomposition depth to $J=2$ brings substantial improvement at mid-resolution (CC* $\approx 0.95$ at 1.9~\AA) but data quality still degrades noticeably at the highest-resolution shell. At $J=3$, CC* remains above 0.91 out to 1.8~\AA\ and $R_\mathrm{split}$ drops below 41\% at the same shell, indicating that three levels of decomposition successfully suppress the majority of the smooth water and air scatter background across scales.

The curves for $J=4$ and $J=5$ are nearly indistinguishable at all resolution shells below the highest, demonstrating that the DWT segmentation has converged: additional decomposition levels do not extract further background. Both reach CC* $> 0.99$ at low-to-mid resolution and $> 0.86$ at 1.74~\AA. The small differences between $J=4$ and $J=5$ in the outermost resolution shell (1.6~\AA) are attributable to limited reflection statistics rather than a systematic quality difference.

Comparing to the baseline, levels 4 and 5 track the baseline closely at all but the highest-resolution shell, confirming that four decomposition levels are sufficient for this dataset and that further levels provide no benefit. Based on these results, $J=4$ is the recommended decomposition depth for this detector geometry and experimental configuration.

\begin{table}[h!]
\centering
\begin{tabular}{|c|ccccc|ccccc|}
\hline
& \multicolumn{5}{c|}{CC*} & \multicolumn{5}{c|}{$R_\mathrm{split}$ (\%)} \\
$d$ (\AA) & $J$=1 & $J$=2 & $J$=3 & $J$=4 & $J$=5 & $J$=1 & $J$=2 & $J$=3 & $J$=4 & $J$=5 \\
\hline
6.1 & 0.973 & 0.985 & 0.989 & 0.991 & 0.990 & 21.7 & 15.1 & 11.9 & 11.1 & 11.4 \\
3.0 & 0.971 & 0.982 & 0.989 & 0.990 & 0.986 & 23.8 & 17.6 & 12.8 & 12.7 & 14.2 \\
2.5 & 0.964 & 0.981 & 0.987 & 0.988 & 0.991 & 25.0 & 16.9 & 12.5 & 12.4 & 11.6 \\
2.1 & 0.897 & 0.971 & 0.973 & 0.981 & 0.983 & 41.0 & 22.1 & 17.7 & 16.0 & 15.8 \\
1.8 & 0.671 & 0.847 & 0.917 & 0.929 & 0.929 & 93.8 & 52.9 & 40.4 & 37.2 & 36.3 \\
1.7 & 0.370 & 0.662 & 0.810 & 0.865 & 0.874 & 138  & 75.6 & 63.5 & 56.5 & 54.0 \\
1.6 & 0.209 & 0.088 & 0.240 & 0.085 & 0.235 & 210  & 430  & 471  & 259  & 251  \\
\hline
\end{tabular}
\caption{CC* and $R_\mathrm{split}$ at selected resolution shells for decomposition levels $J = 1$--$5$. The outermost shell (1.6~\AA) has low reflection counts and is statistically unreliable; practical resolution limit is $\sim$1.7--1.8~\AA\ for levels $\geq 3$.}
\label{tab:ccstar_rsplit_summary}
\end{table}

\section{Hardware Implementation}
\label{sec:hw_implementation}

Having established the DWT segmentation algorithm and validated its crystallographic fidelity in the preceding sections, we now address the practical question of real-time implementation. The frame rates of current and next-generation LCLS-II detectors demand that any on-detector data reduction be realized in dedicated hardware (either FPGA firmware or custom ASIC logic). This section describes the mapping of the DWT2D filters onto FPGA resources, reports place-and-route results on an AMD/Xilinx Alveo U200 platform, and projects the resource footprint for integration into the ePixUHR detector firmware and the upcoming SparkPix readout ASIC.

\subsection{Pre-processing Requirements and Detector Architecture}

The ePixUHR detector currently operates at 35,000~fps; its planned upgrade will reach 100,000~fps. At these rates, all image processing must keep pace with the data stream, making hardware-accelerated algorithms essential for a full data reduction pipeline. The DWT segmentation proposed in Section~\ref{sec:dwt_segmentation} requires that input images have dark offsets removed and, ideally, gain corrections applied so that all pixels present a uniform response. We assume these pre-processing steps are already performed in the data path using existing ASIC and FPGA capabilities (the ePixUHR$_{35,000\text{fps}}$ firmware and SparkPix-RT ASIC both support dark subtraction and gain correction).

The ePixUHR ASIC outputs its $192 \times 168$~pixel image as eight parallel streams, each comprising 24~columns. A natural parallelization strategy is to compute the DWT independently on each partition. To avoid boundary artifacts, overlapping columns from neighboring partitions must be included: for a $7 \times 7$ kernel, three additional columns per side are required, yielding a $(30, 168)$ input tile per core. For a $5 \times 5$ kernel the tile reduces to $(28, 168)$.

\subsection{Mapping DWT2D to FPGA}

Two implementation strategies were evaluated. A separable-kernel approach applies two back-to-back 1D filters but requires buffering the entire image between passes. A 2D convolution (conv2D) approach stores only $N$~lines of the image at a time, reducing Block RAM (BRAM) usage at the cost of higher DSP utilization ($N \times N$ multiplications per clock cycle versus $N$ for the separable case). The key insight enabling both mappings is that the conv2D kernel weights are set to the DWT analysis filter coefficients ($h[n]$ for the low-pass channel and $g[n]$ for the high-pass channel, as defined in Section~\ref{sec:dwt}), with a stride of two implementing the dyadic downsampling; no training is required and the weights are fixed constants. The DWT is thus implemented as a conv2D layer with four output channels (LL, LH, HL, HH) and no bias term, developed using the SLAC Neural Network Library (SNL)~\cite{herbst2022implementation}.

\subsection{FPGA Resource Utilization}

The High-Level Synthesis (HLS) IP was placed and routed on an Alveo U200 at 200~MHz. Table~\ref{tab:fpga performance} reports the resource utilization for several kernel sizes, data types, and initiation intervals (II, the number of clock cycles between successive input samples).

The critical constraint is DSP slice usage. The U200 provides 6,840~DSP slices. A single decomposition layer requires 8~cores per ASIC across 6~ASICs, totaling 48~parallel cores. This limits each core to at most $6{,}840 / 48 = 142$~DSP slices but in practice far fewer, since the design must accommodate all four decomposition layers. Successive layers process $4\times$, $16\times$, and $64\times$ less data, allowing the II to be relaxed to enable DSP reuse while still meeting throughput requirements.

The conv2D-based configurations (upper rows of Table~\ref{tab:fpga performance}) exhaust DSP resources and cannot support a full 4-layer decomposition. The separable-kernel configurations (bottom rows) are substantially more efficient. Estimating the DSP usage for layers 2--4 at II~$= 4$, the total requirement is 69~DSPs for four layers per core, or 3,312~DSPs for the full 48-core design which is roughly 48\% of the U200 budget.

\subsection{Memory Considerations}

In the layer-by-layer pipelined architecture, the LH, HL, and HH subbands must be buffered while the LL subband is forwarded to the next decomposition level. A 4-layer design therefore requires storage for three full images of subband data. Each BRAM block on the Alveo U200 provides 36~kb. The estimated memory requirement is approximately 260~BRAM blocks, well within the 1,766 available on the device.


These resource estimates (48\% DSP utilization and 15\% BRAM occupancy) confirm that a complete 4-level DWT segmentation pipeline is feasible on the Alveo U200 using separable filters at float16 precision. The modest resource footprint is compatible with integration into the ePixUHR FPGA firmware and provides a direct path toward its implementation in the SparkPix readout ASIC family.

\begin{table}[h!]
\centering
\begin{tabular}{llrrrrrrl}
\toprule
Kernel size & Data type & LUT & FF & DSP & BRAM & II & Latency ($\mu$s) & WNS (ns) \\
\midrule
\multicolumn{9}{l}{\textit{Conv2D}} \\
$7 \times 7$         & float32 & 74,846 & 107,160 & 1,072 & 4  & 1 & 26.9 & 0.043 \\
$7 \times 7$         & float16 & 40,754 &  57,713 &   876 & 4  & 1 & 27.2 & 0.387 \\
$5 \times 5$         & float32 & 38,373 &  55,738 &   548 & 4  & 1 & 24.5 & 0.043 \\
$5 \times 5$         & float16 & 20,737 &  30,029 &   448 & 4  & 1 & 24.7 & 0.677 \\
$3 \times 3$         & float32 & 14,756 &  21,033 &   200 & 4  & 1 & 22.3 & 0.096 \\
$3 \times 3$         & float16 &  7,797 &  11,421 &   164 & 4  & 1 & 22.4 & 1.109 \\
\midrule
\multicolumn{9}{l}{\textit{Conv2D low-DSP}} \\
$7 \times 7$         & float16 & 53,983 &  55,494 &   438 & 4  & 1 & 27.1 & 0.030 \\
$5 \times 5$         & float16 & 27,508 &  28,917 &   224 & 4  & 1 & 24.6 & 0.086 \\
$5 \times 5$         & float16 & 19,311 &  22,256 &   112 & 4  & 2 & 48.1 & 0.139 \\
$5 \times 5$         & float16 & 14,335 &  17,878 &    76 & 4  & 3 & 71.6 & 0.107 \\
$5 \times 5$         & float16 & 12,617 &  16,521 &    56 & 4  & 4 & 95.1 & 0.060 \\
\midrule
\multicolumn{9}{l}{\textit{Separable low-DSP}} \\
$5 \times 5$         & float16 &  8,566 &  10,398 &    33 & 12 & 1 & 26.2 & 0.251 \\
$5 \times 5$         & float16 &  4,649 &   6,016 &    12 & 12 & 4 & 24.0 & 0.331 \\
\bottomrule
\end{tabular}
\caption{Single DWT layer (4 HLS cores) resource utilization and performance on an Alveo U200 at 200~MHz. II denotes the initiation interval in clock cycles.}
\label{tab:fpga performance}
\end{table}

\section{Summary}
\label{sec:Summary}

We have presented a 2D discrete wavelet transform (DWT) pipeline for real-time segmentation of background and diffraction signals in serial crystallography images, motivated by the data-rate challenges of LCLS-II/II-HE detectors operating at up to 100,000~fps.

The method was validated on two real LCLS data sets using two distinct detectors (CSPAD and ePix10k) and on 100 simulated nanoBragg frames with known ground truth. A systematic comparison of 12~wavelet families showed that the Haar wavelet (db1) is the best candidate among the tested wavelets for Bragg-peak detection, achieving $F1 \approx 0.96$ at four decomposition levels ($J = 4$), while all longer-filter wavelets clustered at $F1 \approx 0.6$--$0.73$. This advantage stems from Haar's minimal 2-tap support, which preserves the sharp, localized structure of diffraction peaks. At the peak-detection level, DWT2D with Haar substantially outperformed the established peakfinder8 algorithm ($F1 \approx 0.96$ vs.\ $0.37$), with both higher precision ($P \approx 1.00$ vs.\ $0.94$) and dramatically higher recall ($R \approx 0.92$ vs.\ $0.24$). Downstream crystallographic analysis performed on real images from the ePix10kA detector confirmed that the segmentation preserves data quality: CC* and $R_\mathrm{split}$ converge at $J = 4$ and track the unprocessed baseline through the practical resolution limit.

A noise robustness study revealed a limitation of the current pipeline: because LL-zeroing removes background but does not denoise, DWT2D precision degrades under added Gaussian noise exceeding $\sim$50~photons, while peakfinder8 remains stable due to its adaptive radial noise estimation. However, this degradation threshold is well above the expected noise floor of the ePixUHR detector (less than 1~photon), so the method remains robust under realistic operating conditions.

On the hardware side, we evaluated both conv2D and separable-kernel FPGA implementations of the DWT analysis filters on an Alveo U200 at 200~MHz. The DWT filter coefficients are fixed constants that map directly to convolution kernel weights with stride-2 downsampling, requiring no training. While conv2D-based designs exhaust DSP resources, a separable 5$\times$5 low-DSP design requires only 69~DSP slices per core for a full 4-layer decomposition, projecting to $\sim$48\% of total U200 DSP resources for the complete 48-core, 6-ASIC design, which is within budget for integration into the ePix~UHR FPGA firmware, with a natural path to on-detector ASIC integration in future SparkPix generations.

The proposed algorithm and the demonstration of the convolution layer in hardware present a concept of how data segmentation and analyses can be performed in upcoming high speed detectors to enable data reduction in real time. These concepts are introduced at the FPGA level but can be extended to ASIC implementation. Full implementation of these structures was beyond the scope of this work and is considered future work.

\acknowledgments
R\&D at the Linac Coherent Light Source (LCLS), SLAC National Accelerator Laboratory, is supported by the U.S. Department of Energy, Office of Science, Office of Basic Energy Sciences under Contract No. DE-AC02-76SF00515. AI assisted technology was used in the preparation of this manuscript.

\bibliographystyle{JHEP}  
\bibliography{bibliography}   

@article{Kieffer2025,
  author  = {Kieffer, J\'er\^ome and Orlans, Julien and Coquelle, Nicolas and Debionne, Samuel
             and Basu, Shibom and Homs, Alejandro and Santoni, Gianluca and De Sanctis, Daniele},
  title   = {Application of signal separation to diffraction image compression and serial crystallography},
  journal = {Journal of Applied Crystallography},
  year    = {2025},
  volume  = {58},
  pages   = {138--153},
  doi     = {10.1107/S1600576724011038},
}

@article{HadianJazi2021,
  author  = {Hadian-Jazi, Marjan and Sadri, Alireza and Barty, Anton and Yefanov, Oleksandr
             and Galchenkova, Marina and Oberthuer, Dominik and Komadina, Dana and Brehm, Wolfgang
             and Kirkwood, Henry and Mills, Grant and de Wijn, Raphael and Letrun, Romain
             and Kloos, Marco and Vakili, Mohammad and Gelisio, Luca and Darmanin, Connie
             and Mancuso, Adrian P. and Chapman, Henry N. and Abbey, Brian},
  title   = {Data reduction for serial crystallography using a robust peak finder},
  journal = {Journal of Applied Crystallography},
  year    = {2021},
  volume  = {54},
  pages   = {1360--1378},
  doi     = {10.1107/S1600576721007317},
}

@inproceedings{herbst2022implementation,
  title={Implementation of a framework for deploying AI inference engines in FPGAs},
  author={Herbst, Ryan and Coffee, Ryan and Fronk, Nathan and Kim, Kukhee and Kim, Kuktae and Ruckman, Larry and Russell, JJ},
  booktitle={Smoky Mountains Computational Sciences and Engineering Conference},
  pages={120--134},
  year={2022},
  organization={Springer}
}

@article{Starck1994,
  author  = {Starck, J.-L. and Murtagh, F.},
  title   = {Image restoration with noise suppression using the wavelet transform},
  journal = {Astronomy \& Astrophysics},
  year    = {1994},
  volume  = {288},
  pages   = {343--348},
}

@article{rasheedi202528,
title={A 28 nm multiply-accumulate ASIC architecture for on-chip data compression in MHz frame rate X-ray and electron pixel detectors},
author={Rasheedi, Rami and Contini, Nicholas and Gharib, Mohamed Adel and Strempfer, Sebastian and Gnanasekaran, Senthil and Abdelzaher, Salma and Guruswamy, Tejas and Yoshii, Kazutomo and Hammer, Mike and Shi, Henry and others},
journal={Journal of Instrumentation},
volume={20},
number={10},
pages={P10027},
year={2025},
publisher={IOP Publishing},
doi = {10.1088/1748-0221/20/10/P10027},
url = {https://doi.org/10.1088/1748-0221/20/10/P10027},
}

@inproceedings{king2023characterization,
  title={Characterization of ePixUHR-35kHz: a 13.8 Gb/s full-frame X-ray imaging readout ASIC for the LCLS-II upgrade},
  author={King, P and Doering, D and Gupta, A and Habib, A and Hansson, C and Hammer, MP and Kenney, C and Markovic, B and Miceli, A and Pena-Perez, A and others},
  booktitle={2023 IEEE Nuclear Science Symposium, Medical Imaging Conference and International Symposium on Room-Temperature Semiconductor Detectors (NSS MIC RTSD)},
  pages={1--1},
  year={2023},
  organization={IEEE}
}

@article{rota2024sparkpix,
  title={The SparkPix-S ASIC for the sparsified readout of 1 MHz frame-rate X-ray cameras at LCLS-II: pixel design and simulation results},
  author={Rota, L and Mele, Fillippo and Habib, A and Kim, H and King, P and Markovic, B and Perez, A Pe{\~n}a and Dragone, A},
  journal={Journal of Instrumentation},
  volume={19},
  number={01},
  pages={C01010},
  year={2024},
  publisher={IOP Publishing}
}

@inproceedings{markovic2023sparkpix,
  title={SparkPix-T: Spatial and Time Resolving Front-End ASIC with MHz-Rate Information Extraction for Momentum Spectroscopy at LCLS-II},
  author={Markovic, B and Bakalis, C and Blaj, G and Defay, X and Doering, D and Gupta, A and Hasi, J and Kenney, C and King, P and Pena-Perez, A and others},
  booktitle={2023 IEEE Nuclear Science Symposium, Medical Imaging Conference and International Symposium on Room-Temperature Semiconductor Detectors (NSS MIC RTSD)},
  pages={1--1},
  year={2023},
  organization={IEEE}
}

@techreport{schoenlein2016lcls,
  title={LCLS-II High Energy (LCLS-II-HE): a transformative X-ray laser for science},
  author={Schoenlein, Robert W and Adolphsen, C and Mori, R Alonso and Aquila, A and Bare, S and Bargar, J and Bergmann, U and Boutet, S and Brown, G and Bucksbaum, P and others},
  year={2016},
  institution={SLAC National Accelerator Laboratory (SLAC), Menlo Park, CA (United States)}
}

@article{Sandberg2025,
	title = {{GT} {Readout} — {A} development platform for 1 {MHz} frame-rate detectors at {LCLS}-{II}},
	volume = {20},
	url = {https://doi.org/10.1088/1748-0221/20/08/P08019},
	doi = {10.1088/1748-0221/20/08/P08019},
	number = {08},
	journal = {Journal of Instrumentation},
	author = {Sandberg, H. and King, P. and Doering, D. and Valle, J. and Oriunno, M. and Rota, L. and Dragone, A.},
	month = aug,
	year = {2025},
	note = {Publisher: IOP Publishing},
	pages = {P08019},
}

@article{Parkhurst2014_dxtbx,
author = "Parkhurst, James M. and Brewster, Aaron S. and Fuentes-Montero, Luis and Waterman, David G. and Hattne, Johan and Ashton, Alun W. and Echols, Nathaniel and Evans, Gwyndaf and Sauter, Nicholas K. and Winter, Graeme",
title = "{{\it dxtbx}: the diffraction experiment toolbox}",
journal = "Journal of Applied Crystallography",
year = "2014",
volume = "47",
number = "4",
pages = "1459--1465",
month = "Aug",
doi = {10.1107/S1600576714011996},
url = {https://doi.org/10.1107/S1600576714011996},
}

@article{Mendez2024_DeepResidualNets,
author = "Mendez, Derek and Holton, James M. and Lyubimov, Artem Y. and Hollatz, Sabine and Mathews, Irimpan I. and Cichosz, Aleksander and Martirosyan, Vardan and Zeng, Teo and Stofer, Ryan and Liu, Ruobin and Song, Jinhu and McPhillips, Scott and Soltis, Mike and Cohen, Aina E.",
title = "{Deep residual networks for crystallography trained on synthetic data}",
journal = "Acta Crystallographica Section D",
year = "2024",
volume = "80",
number = "1",
pages = "26--43",
month = "Jan",
doi = {10.1107/S2059798323010586},
url = {https://doi.org/10.1107/S2059798323010586},
}

@incollection{Young2023_Chp7,
title = {Chapter Seven - Interpreting macromolecular diffraction through simulation},
editor = {Nozomi Ando},
series = {Methods in Enzymology},
publisher = {Academic Press},
volume = {688},
pages = {195-222},
year = {2023},
booktitle = {Crystallography of Protein Dynamics},
issn = {0076-6879},
doi = {https://doi.org/10.1016/bs.mie.2023.06.011},
url = {https://www.sciencedirect.com/science/article/pii/S0076687923002161},
author = {Iris D. Young and Derek Mendez and Billy K. Poon and Johannes P. Blaschke and Felix Wittwer and Michael E. Wall and Nicholas K. Sauter},
}

@article{Barty2014,
author = {Barty, Anton and Kirian, Richard A. and Maia, Filipe R. N. C. and Hantke, Max F. and Yoon, Chun Hong and White, Thomas A. and Chapman, Henry N.},
title = {Cheetah: software for high-throughput reduction and analysis of serial femtosecond X-ray diffraction data},
journal = {Journal of Applied Crystallography},
volume = {47},
pages = {1118--1131},
year = {2014},
doi = {10.1107/S1600576714007626}
}

@article{White2012,
author = {White, Thomas A. and Kirian, Richard A. and Martin, Adrian V. and Aquila, Andrew and Nass, Karol and Barty, Anton and Chapman, Henry N.},
title = {CrystFEL: a software suite for snapshot serial crystallography},
journal = {Journal of Applied Crystallography},
volume = {45},
pages = {335--341},
year = {2012},
doi = {10.1107/S0021889812002312}
}

@article{White2016,
author = {White, Thomas A.},
title = {Processing serial crystallography data with CrystFEL: a step-by-step guide},
journal = {Acta Crystallographica Section D},
volume = {72},
pages = {1235--1242},
year = {2016},
doi = {10.1107/S205979831801238X}
}

@article{Kabsch2010,
author = {Kabsch, Wolfgang},
title = {XDS},
journal = {Acta Crystallographica Section D},
volume = {66},
pages = {125--132},
year = {2010},
doi = {10.1107/S0907444909047337}
}

@article{Powell1999,
author = {Powell, Harold R.},
title = {The Rossmann Fourier autoindexing algorithm in MOSFLM},
journal = {Acta Crystallographica Section D},
volume = {55},
pages = {1690--1695},
year = {1999},
doi = {10.1107/S0907444999009506}
}

@article{Karplus2012,
author = {Karplus, P. A. and Diederichs, K.},
title = {Linking crystallographic model and data quality},
journal = {Science},
volume = {336},
pages = {1030--1033},
year = {2012},
doi = {10.1126/science.1218231}
}

@article{Mallat1989,
  author  = {Mallat, St{\'e}phane G.},
  title   = {A theory for multiresolution signal decomposition: the wavelet representation},
  journal = {IEEE Transactions on Pattern Analysis and Machine Intelligence},
  volume  = {11},
  number  = {7},
  pages   = {674--693},
  year    = {1989},
  doi     = {10.1109/34.192463}
}

@article{BraggNN_Liu2022,
  author  = {Liu, Zhengchun and Sharma, Hemant and Park, Jun-Sang and Kenesei, Peter and Almer, Jonathan and Kettimuthu, Rajkumar and Foster, Ian and Yusuf, Syed},
  title   = {{BraggNN}: fast {X}-ray {Bragg} peak analysis using deep learning},
  journal = {IUCrJ},
  volume  = {9},
  number  = {1},
  pages   = {104--113},
  year    = {2022},
  doi     = {10.1107/S2052252521011258}
}

@article{ke2018convolutional,
  title={A convolutional neural network-based screening tool for X-ray serial crystallography},
  author={Ke, T-W and Brewster, Aaron S and Yu, Stella X and Ushizima, Daniela and Yang, Chao and Sauter, Nicholas K},
  journal={Synchrotron Radiation},
  volume={25},
  number={3},
  pages={655--670},
  year={2018},
  publisher={International Union of Crystallography}
}

@article{PeakNet_Peck2025,
  author  = {Peck, Aidan and Donatelli, Jeffrey J. and Brewster, Aaron S. and Sauter, Nicholas K.},
  title   = {{PeakNet}: a deep learning approach to {Bragg} peak detection in serial crystallography},
  journal = {Frontiers in High Performance Computing},
  volume  = {3},
  pages   = {1545943},
  year    = {2025},
  doi     = {10.3389/fhpcp.2025.1545943}
}

@misc{psalgos,
  author = {{SLAC National Accelerator Laboratory}},
  title = {psalgos: Peak Finding Algorithms for LCLS Data},
  year = {2023},
  howpublished = {\url{https://lcls-psana.github.io/psalgos/}},
  note = {Accessed: 2026-05-03}
}

@article{BusingLevy1967,
  author    = {Busing, W. R. and Levy, H. A.},
  title     = {Angle Calculations for 3- and 4-Circle X-ray and Neutron Diffractometers},
  journal   = {Acta Crystallographica},
  volume    = {22},
  number    = {4},
  pages     = {457--464},
  year      = {1967},
  doi       = {10.1107/S0365110X67000970}
}

@article{BrehmDiederichs2014,
  author  = {Brehm, Wolfgang and Diederichs, Kay},
  title   = {Breaking the indexing ambiguity in serial crystallography},
  journal = {Acta Crystallographica Section D: Biological Crystallography},
  volume  = {70},
  number  = {1},
  pages   = {101--109},
  year    = {2014},
  doi     = {10.1107/S1399004713025431}
}

\end{document}